\RequirePackage[2020-02-02]{latexrelease}

\documentclass[twocolumn,preprintnumbers]{revtex4}%
\usepackage{amssymb}
\usepackage{amsmath}
\usepackage{graphicx}
\usepackage{epstopdf}
\usepackage{dcolumn}
\usepackage{bm}
\usepackage{xcolor}
\usepackage{ifthen}
\usepackage[normalem]{ulem}
\usepackage{amsfonts}%
\RequirePackage[2020-02-02]{latexrelease}
\graphicspath{{c:/Users/Eyal/MyDocuments/latex/Texfiles/Diamond/dipolarBA/EntangSupp/SDIP/}
	{c:/Users/Eyal/MyDocuments/latex/Texfiles/Diamond/dipolarBA/EntangSupp/SDD/}
	{c:/Users/Eyal/MyDocuments/latex/Texfiles/Diamond/dipolarBA/EntangSupp/SDH/}
	{c:/Users/eyal/Documents/latex/SDD/}
	{c:/Users/Eyal/MyDocuments/latex/Texfiles/Diamond/fMixYIG/}
	{c:/Users/Eyal/MyDocuments/latex/Texfiles/Diamond/FerriMO/YIGS/}}
\UseRawInputEncoding
\providecommand{\U}[1]{\protect\rule{.1in}{.1in}}
\providecommand{\U}[1]{\protect\rule{.1in}{.1in}}
\def\showal{1}
\newcommand{\al}[1]{\ifthenelse{\showal=1}{\textcolor{orange}{[[#1]]}}{}}

\newcommand{\eb}[1]{\ifthenelse{\showal=1}{\textcolor{cyan}{[[#1]]}}{}}
\begin{document}
\title{Parametric excitation of a ferrimagnetic sphere resonator}
\author{Eyal Buks}
\email[]{eyal@ee.technion.ac.il}
\affiliation{Andrew and Erna Viterbi Department of Electrical Engineering, Technion, Haifa
32000 Israel}
\date{\today }

\begin{abstract}
The response of a ferrimagnetic sphere resonator to an externally applied
parametric excitation is experimentally studied. Measurement results are
compared with predictions derived from a theoretical model, which is based on
the hypothesis that disentanglement spontaneously occurs in quantum systems.
According to this hypothesis, time evolution is governed by a modified master
equation having an added nonlinear term that deterministically generates
disentanglement. It is found that the disentanglement--based model is
compatible with the experimental results. In particular, the model can
qualitatively account for an experimentally observed
instability in the system under study, which cannot be derived from any
theoretical model that is based on a linear master equation.

\end{abstract}
\pacs{}
\maketitle

\textbf{Introduction} -- Multistability is experimentally observed
in many quantum systems. In contrast, the linearity of standard quantum
mechanics (QM) excludes multistability in systems having Hilbert space of finite dimensionality. In the current study, this apparent conflict is experimentally and theoretically explored by
studying the response of a ferrimagnetic sphere resonator (FMSR) to an
externally applied parametric excitation.

In the classical realm, the response of a resonator to an externally applied
parametric excitation (longitudinal driving) is well described by the Mathieu
model \cite{Mathieu_137}. When the excitation frequency is tuned to a
parametric resonance, above a critical value of the parametric excitation
amplitude \cite{Landau1960a} the system's steady state response becomes
bistable \cite{Hioki_203901}. Moreover, with an added applied monochromatic
forcing (transverse driving), the resonator's response exhibits dependency on
the relative phase between longitudinal and transverse driving tones. This
dependency can be exploited for the construction of amplifiers having a
phase--sensitive gain.

In the quantum realm, similar effects can be explored using spins
\cite{Wanic_134257}. Longitudinal driving can be applied by modulating the
magnetic field parallel to the spins' magnetization vector, whereas modulating
the perpendicular component gives rise to transverse driving. The effect of
parametric excitation (i.e. longitudinal driving, aka parallel pumping)
applied to magnetically ordered dielectrics has been extensively studied
\cite{Lvov_233}. Experimental observation of a response qualitatively similar
to what is predicted by the Mathieu model has been first reported in
\cite{Bloembergen_699}. In the region where multistability occurs, the
experimentally observed spins' response becomes hysteretic.

Previously proposed theoretical explanations for experimentally observed
multistabilities in finite quantum systems are based on the assumption that
time evolution is nonlinear. For spin systems, nonlinearity can be introduced
by implementing the Holstein--Primakoff transformation \cite{Holstein_1098},
which allows expressing spin operators in terms of annihilation and creation
Bosonic operators. In this transformation, which is henceforth referred to as
Bosonization, the spins' Hilbert space having finite dimensionality is mapped
into a space having infinite dimensionality. In the presence of magnetic
anisotropy, this method gives rise to nonlinearity in the time evolution,
which, in turn, enables both instability and multistability \cite{Wang_224410}. In contrast, for finite quantum systems, both instability and
multistability are theoretically excluded provided that time evolutions for
the systems's reduced density operator $\rho$ is governed by a master equation
that linearly depends on $\rho$ \cite{Nigro_043202}.

The above--discussed difficulty to justify the Bosonization--based model,
which enables multistabilities that are otherwise theoretically excluded
\cite{Buks_012439}, is the main motivation for the current study. Here, an
alternative theoretical model, which is based on the hypothesis that
disentanglement spontaneously occurs in quantum systems \cite{Buks_2400036},
is explored. According to this hypothesis, time evolution is governed by a
modified master equation having an added nonlinear term [see Eq. (\ref{MME})
below]. A FMSR is used to experimentally validate the proposed model. While the impact of disentanglement on the FMSR response to transverse driving has been explored in \cite{Buks_2400587}, here the effect of parametric excitation is studied. It is found that the disentanglement--based model can qualitatively account for an instability, which is
experimentally--observed in the system under study, and which is arguably
inconsistent with any linear master equation.

\textbf{Modified master equation} -- The spontaneous disentanglement
hypothesis is based on the assumption that time evolution is governed by a
master equation for the reduced density operator $\rho$ having a form given by
\cite{Grimaudo_033835,Buks_2400036,Kowalski_167955,Elben_200501,Beretta_026113}%
\begin{equation}
\frac{\mathrm{d}\rho}{\mathrm{d}t}=i\hbar^{-1}\left[  \rho,\mathcal{H}\right]
+\mathcal{L}-\Theta\rho-\rho\Theta+2\left\langle \Theta\right\rangle \rho\;,
\label{MME}%
\end{equation}
where $\hbar$ is the Planck's reduced constant, $\mathcal{H}^{{}}=\mathcal{H}^{\dag}$
is the Hamiltonian, $\mathcal{L}$ is a Lindblad superoperator
\cite{Lindblad_119} (which linearly depends on $\rho$), $\Theta^{{}}%
=\Theta^{\dag}$ and $\left\langle \Theta\right\rangle =\operatorname{Tr}%
\left(  \Theta\rho\right)  $. The added term $-\Theta\rho-\rho\Theta
+2\left\langle \Theta\right\rangle \rho$ in Eq. (\ref{MME}) deterministically
generates disentanglement. The dependency of the disentanglement operator
$\Theta$ on $\rho$ gives rise to nonlinear dynamics. The construction of the
disentanglement operator $\Theta$ is explained in \cite{Buks_2400036}. The
disentanglement process is characterized by a rate denoted by $\gamma
_{\mathrm{D}}$. For spin systems, the Lindblad superoperator $\mathcal{L}$ is
characterized by energy--relaxation $\Gamma_{1}$ and dephasing $\Gamma
_{\varphi}$ rates, thermal occupation factor $\hat{n}_{0}$, and longitudinal
$T_{1}$ and transverse $T_{2}$ relaxation times [see Eq. (17.154) of Ref.
\cite{Buks_QMLN}].

\textbf{Driven $L$ spin system} -- The system under study is composed of $L$
coupled spins 1/2. The total angular momentum vector operator $\mathbf{S}%
=\left(  S_{x},S_{y},S_{z}\right)  $ in units of $\hbar/2$, is given by
$\mathbf{S}=\sum_{l=1}^{L}\mathbf{S}_{l}$, where $\mathbf{S}_{l}=\left(
S_{l,x},S_{l,y},S_{l,z}\right)  $ is the $l$'th spin angular momentum vector
operator. The closed-system Hamiltonian $\mathcal{H}$ is given by \cite{Buks_2400587}%
\begin{align}
\frac{\mathcal{H}}{\hbar}  &  =-\frac{\omega_{z}S_{z}}{2}\nonumber\\
&  +\frac{\omega_{\mathrm{K}}\left(  S_{+}S_{-}+S_{-}S_{+}\right)
+\omega_{\mathrm{A}}\left(  S_{+}^{2}+S_{-}^{2}\right)  }{8}\nonumber\\
&  +\frac{S_{+}\Omega_{\mathrm{T}1}^{{}}e^{i\omega_{\mathrm{T}}t}+S_{-}%
\Omega_{\mathrm{T}1}^{\ast}e^{-i\omega_{\mathrm{T}}t}}{4}\;,\nonumber\\
&  \label{H DLS}%
\end{align}
where $S_{\pm}=S_{x}\pm iS_{y}$, the rates $\omega_{\mathrm{K}}$,
$\omega_{\mathrm{A}}$, $\omega_{\mathrm{T}}=\omega_{0}+\omega_{\mathrm{d}}$
(transverse driving angular frequency), $\omega_{0}$ (angular resonance
frequency) and $\omega_{\mathrm{d}}$ (transverse driving angular detuning
frequency)\ are real constants, $\Omega_{\mathrm{T}1}=\left\vert
\Omega_{\mathrm{T}1}\right\vert e^{i\phi_{\mathrm{T}}}$ (transverse driving
amplitude) is a complex constant, the real time--dependent angular frequency
$\omega_{z}$\ is given by $\omega_{z}=\omega_{0}+\Omega_{\mathrm{L}1}^{{}}%
\cos\left(  2\left(  \omega_{\mathrm{T}}+\omega_{\mathrm{f}}\right)  t\right)
$, where $\omega_{\mathrm{f}}$ (longitudinal driving angular detuning
frequency) and $\Omega_{\mathrm{L}1}^{{}}$ (longitudinal driving amplitude)
are real constants. The terms $S_{+}S_{-}+S_{-}S_{+}=2\left(  S_{x}^{2}%
+S_{y}^{2}\right)  $ and $S_{+}^{2}+S_{-}^{2}=2\left(  S_{x}^{2}-S_{y}%
^{2}\right)  $ in the Hamiltonian $\mathcal{H}$ (\ref{H DLS})\ account for
magnetic anisotropy. The following commutation relations hold $\left[
S_{i},S_{j}\right]  =2i\epsilon_{ijk}S_{k}$, $\left[  S_{z},S_{\pm}\right]
=\pm2S_{\pm}$, and $\left[  S_{+},S_{-}\right]  =4S_{z}$. Equations of motion
obtained from the Hamiltonian $\mathcal{H}$ (\ref{H DLS}) (without
disentanglement) are derived in section \ref{SM_EOM} of the supplementary
materials (SM). Note that alternative methods to model dipolar
coupling are reviewed in \cite{Stancil_Spin}.

\begin{figure}[ptb]
\begin{center}
\includegraphics[width=3.0in,keepaspectratio]{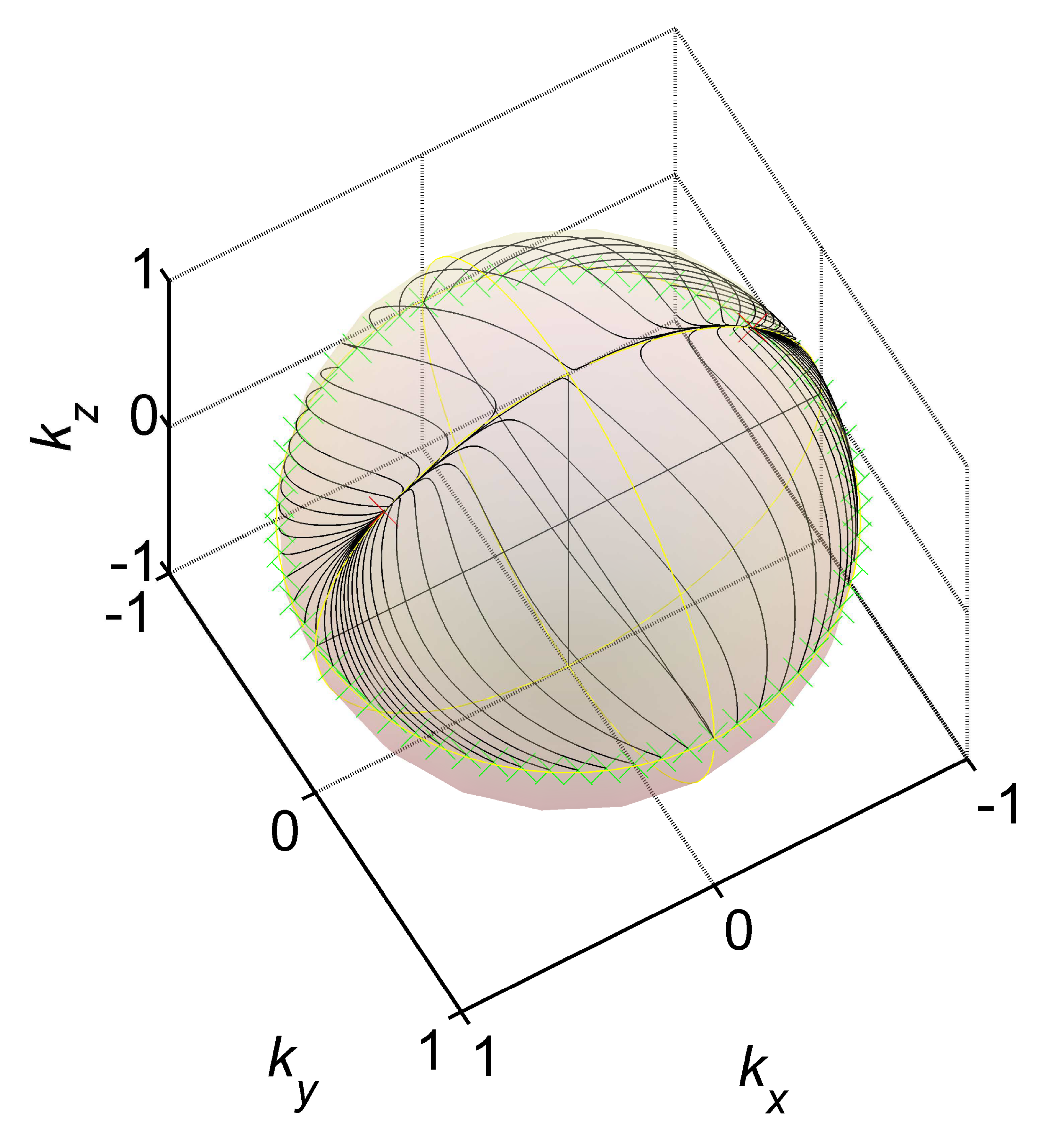}
\end{center}
\caption{{} Two spins. Time evolution of the
magnetization $\left\langle \mathbf{S}\right\rangle =\left(  k_{x},k_{y}%
,k_{z}\right)  $ for the case $L=2$. Initial states along the Bloch sphere
equator are labelled by green $\times$ symbols, and the two locally--stable
steady states by red $\times$ symbols. Assumed parameters' values are
$\omega_{\mathrm{K}}=0$, $\Omega_{\mathrm{T}1}=0$, $\omega_{\mathrm{d}}=0$,
$\gamma_{\mathrm{D}}/\omega_{\mathrm{A}}=100$, $\Omega_{\mathrm{L}1}%
/\omega_{\mathrm{A}}=3/2$ and $\omega_{\mathrm{A}} T_{1}=\omega_{\mathrm{A}}
T_{2}=0.2$.}%
\label{FigTS}%
\end{figure}

\textbf{The two--spin case} -- For sufficiently small number $L$ of spins, the
effect of disentanglement can be numerically explored. Matrix representation
of the Hamiltonian $\mathcal{H}$ [see Eq. (\ref{H DLS})] for the case $L=2$
(i.e. two spins ) is derived in SM section \ref{SM_L2}. Time evolution of the
magnetization $\left\langle \mathbf{S}\right\rangle $, which is derived by
numerically integrating the modified master equation (\ref{MME}), is shown in
the plot in Fig. \ref{FigTS}. For the assumed parameters' values (which are
listed in the figure caption), the Bloch sphere is divided into two basins of
attraction corresponding to two locally--stable steady state solutions of the
modified master equation (see the red $\times$ symbols in Fig. \ref{FigTS}).
For these two steady states $\phi_{2}-\phi_{1}=\pi$, where $\phi_{n}$ is the
oscillation phase of the $n$'th steady state with respect to the externally
applied parallel pumping, and where $n\in\left\{  1,2\right\}  $.

This bistability, which is induced by disentanglement, resembles the
above--threshold response of a parametrically driven classical resonator.
However, while the magnetization $\left\langle \mathbf{S}\right\rangle $ is
bounded inside the Bloch sphere in the disentanglement--based model (for any
finite number of spins $L$), the amplitude of a classical resonator is unbounded.

\textbf{Rapid disentanglement limit} -- The demonstration shown in Fig.
\ref{FigTS} for disentanglement--induced bistability of a two--spin system
(i.e. $L=2$) is performed by numerically integrating the master equation
(\ref{MME}). This method becomes intractable unless $L$ is kept sufficiently
small. On the other hand, the rapid disentanglement approximation, which is
based on the assumption that the rate of disentanglement $\gamma_{\mathrm{D}}$
is much larger than all decay rates, can significantly simplify the dynamics.

A theoretical model based on the rapid disentanglement approximation \cite{Buks_2400587} is
derived in SM section \ref{SM_RDM}. It is found that in this approximation,
the expectation value $P_{z}=\left\langle S_{z}\right\rangle $ satisfies in
steady state a cubic polynomial equation given by [see Eq. (\ref{z= DLS}) in
SM section \ref{SM_RDM}]%
\begin{equation}
\frac{P_{z}}{P_{z0}}=\frac{1}{1+\frac{2W}{\alpha+\left(  \delta-4\sqrt{D}%
\frac{P_{z}}{P_{z0}}\right)  ^{2}}}\;, \label{P_z RD SS}%
\end{equation}
where $P_{z0}$ represents the steady state value of $P_{z}$ for the case where
$\Omega_{\mathrm{T}1}^{{}}=0$ and $\Omega_{\mathrm{L}1}^{{}}=0$ (no driving),
$W$ is a phase--dependent\ dimensionless driving amplitude, $\delta$ is a
dimensionless detuning, and $\sqrt{D}$ is a dimensionless transverse lifetime.
To first order in the ratio $\omega_{\mathrm{A}}/\omega_{\mathrm{K}}$, the
dimensionless parameters $W$, $\delta$ and $D$ are given by $W=\left(
1/2\right)  \left\vert \Omega_{\mathrm{T}1}^{{}}\right\vert ^{2}T_{1}%
T_{2}\left[  1+\sqrt{1-\alpha}\sin\left(  2\phi_{\mathrm{T}}\right)  \right]
$, $\delta=\left(  \omega_{\mathrm{T}}-\omega_{0}\right)  T_{2}$ and $\sqrt
{D}=\omega_{\mathrm{K}}T_{2}P_{z0}/4$, and to lowest nonvanishing order in
$\omega_{\mathrm{A}}/\omega_{\mathrm{K}}$, $\alpha$ is given by $\alpha
=1-\left(  \left(  1/2\right)  \left(  \omega_{\mathrm{A}}/\omega_{\mathrm{K}%
}\right)  \Omega_{\mathrm{L}1}^{{}}T_{2}\right)  ^{2}$.

Bistability occurs in the region where the cubic polynomial equation
(\ref{P_z RD SS}) has three real solutions (two of which representing locally
stable steady states). For the case $\alpha\geq D$ bistability is excluded.
For $0<\alpha<D$, bistability is possible for dimensionless driving amplitudes
$W$ bounded by $W\in\left(  W_{-},W_{+}\right)  $ (see Fig. \ref{FigRDz} in SM
section \ref{SM_RDM}). Analytical expressions for the lower $W_{-}$ and upper
$W_{+}$ bounds are derived in SM section \ref{SM_RDM}.

\textbf{Experimental setup} -- A FMSR is employed for experimentally testing
predictions derived from the spontaneous disentanglement hypothesis. This
magnetically--tunable spin system \cite{Stancil_Spin} has a variety of
applications in many fields, including magnonics
\cite{Suhl_209,Zheng_151101,Rameshti_1,Kusminskiy_299,Wang_057202,Wang_224410,Hyde_174423,Juraschek_094407}
and quantum data processing
\cite{Lachance_070101,Lachance_1910_09096,Tabuchi_729,Elyasi_054402,Zhang_023021}%
.

\begin{figure}[ptb]
\begin{center}
\includegraphics[width=3.0in,keepaspectratio]{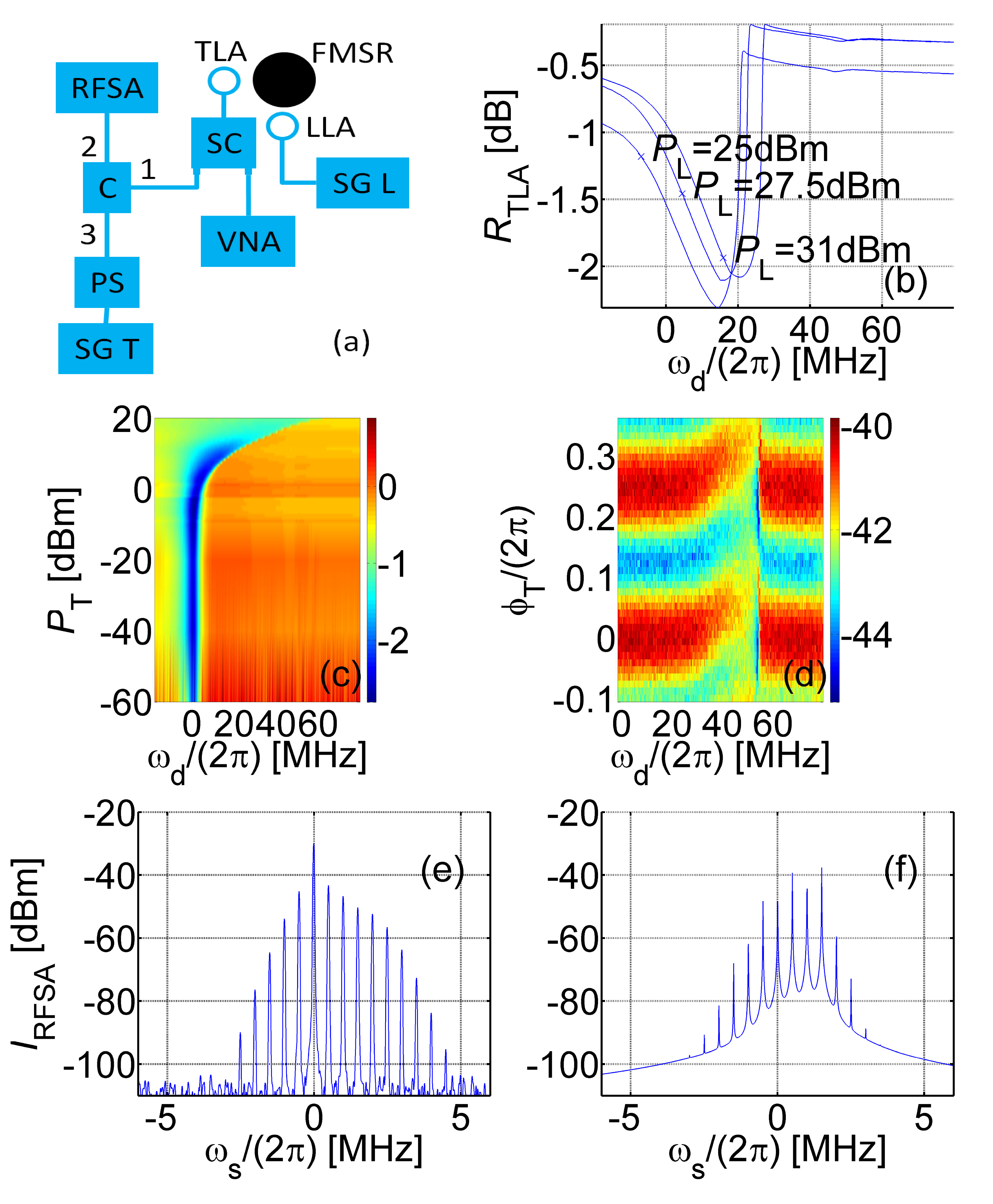}
\end{center}
\caption{{} FMSR. (a) Experimental setup. The FMSR is inductively coupled to two loop antennas (LA), which allow both
driving and detection of FMSR magnetic resonance. The signal generators
labeled as SG L and SG T drive the longitudinal and transverse loop antennas
(LLA and TLA), respectively. A tunable phase shifter (PS) controls the
relative phase $\phi_{\mathrm{T}}$ between longitudinal and transverse driving
tones [see Eq. (\ref{H DLS})]. A circulator (C) and a splitter/combiner (SC)
are used to direct the input and output microwave signals. The
LLA (TLA) axis is parallel (perpendicular) to the applied static magnetic
field $\mathbf{H}_{\mathrm{s}}$. All measurements are performed at room
temperature. The transverse driving frequency $\omega_{\mathrm{T}}/\left(
2\pi\right)  $ is set to $1.874 \operatorname{GHz}$, and the resonance
frequency $\omega_{0}/\left(  2\pi\right)  $ is tuned by adjusting the
electromagnet current. (b) TLA reflectivity $R_{\mathrm{TLA}}$ as a function
of transverse driving detuning frequency $\omega_{\mathrm{d}}/\left(
2\pi\right)  =\left(  \omega_{\mathrm{T}}-\omega_{0}\right)  /\left(
2\pi\right)  $ for 3 different values of the driving power applied to the LLA,
which is denoted by $P_{\mathrm{L}}$. The transverse driving power, which is
denoted by $P_{\mathrm{T}}$, is $10$ dBm. (c) TLA reflectivity
$R_{\mathrm{TLA}}$ (in dB units) as a function of transverse driving detuning
frequency $\omega_{\mathrm{d}}/\left(  2\pi\right)  $ and $P_{\mathrm{T}}$
(for this measurement no longitudinal driving is applied). (d) Intermodulation
peak intensity (in dBm units) as a function of transverse driving detuning
frequency $\omega_{\mathrm{d}}/\left(  2\pi\right)  $ and relative phase
$\phi_{\mathrm{T}}$ (controlled by the PS). The longitudinal driving detuning
frequency $\omega_{\mathrm{f}}/\left(  2\pi\right)  $ is set to $5
\operatorname{kHz}$. The intermodulation peak at frequency $\left(
\omega_{\mathrm{T}}+2\omega_{\mathrm{f}}\right)  /\left(  2\pi\right)  $ is
measured using the RFSA. For this measurement $P_{\mathrm{T}}=10$ dBm and
$P_{\mathrm{L}}=30$ dBm. (e) Mixing with low--frequency longitudinal driving.
The measured RFSA intensity $I_{\mathrm{RFSA}}$ is plotted as a function of
$\omega_{\mathrm{s}} \equiv\omega_{\mathrm{RFSA}} - \omega_{\mathrm{T}}$,
where $\omega_{\mathrm{RFSA}}$ is the RFSA angular frequency. Longitudinal
driving frequency is $\omega_{\mathrm{L}}/\left(  2\pi\right)
=0.5\operatorname{MHz}$, and power is 0 dBm. (f) Theoretical calculation of
$I_{\mathrm{RFSA}}$ based on Eq. (D11) of Ref. \cite{Buks_033807}. FMSR
measured parameters that are used for the calculation are $\omega_{\mathrm{L}%
}T_{1}=0.6$ and $T_{1}/T_{2}=2.1$.}%
\label{FigESP}%
\end{figure}

A sketch of the experimental setup is shown in Fig. \ref{FigESP}(a). The FMSR,
which has a radius of $R_{\mathrm{s}}=125\operatorname{   \mu m}$, is made of
Calcium Vanadium Bismuth Iron Garnet (CVBIG, $\mathrm{Ca}_{2}\mathrm{V}%
_{}\mathrm{Bi}_{}\mathrm{Fe}_{4}\mathrm{O}_{12}$). The angular frequency of
the FMSR Kittel (uniform) mode $\omega_{0}$ is approximately given by
$\omega_{0}=\mu_{0}\gamma_{\mathrm{e}}H_{\mathrm{s}}$ \cite{Walker_390}, where
$\mathbf{H}_{\mathrm{s}}$ is the static magnetic field, $H_{\mathrm{s}%
}=\left\vert \mathbf{H}_{\mathrm{s}}\right\vert $, $\mu_{0}$ is the free space
permeability, and $\gamma_{\mathrm{e}}/2\pi=28\operatorname{GHz}%
\operatorname{T}^{-1}$ is the gyromagnetic ratio \cite{Fletcher_687}. The
static magnetic field $\mathbf{H}_{\mathrm{s}}$ is applied using an
electromagnet. In comparison with Yttrium-Iron-Garnet
\cite{Serga_264002}, which is more commonly used, the smallest value of
$H_{\mathrm{s}}$, for which the FMSR becomes fully magnetized, is
significantly lower in CVBIG \cite{Lynch_225}. To allow lowering $\omega
_{0}/\left(  2 \pi\right)  $ well below $3 \operatorname{GHz}$, CVBIG was
chosen for the current study. FMSR reflectivity is measured
using a vector network analyzer (VNA), and FMSR response is monitored using a
radio frequency spectrum analyzer (RFSA).

The plots shown in Fig. \ref{FigESP}(b-f) demonstrate some of the well--known
nonlinear effects that are observable with FMSRs \cite{Lvov_233}.
Driving--induced resonance line shape distortion is demonstrated by the plots
in Fig. \ref{FigESP} (b) and (c), which exhibit measurements of TLA
reflectivity $R_{\mathrm{TLA}}$. The dependency of resonance line shape on
longitudinal $P_{\mathrm{L}}$ and transverse $P_{\mathrm{T}}$ driving powers,
that are applied to the LLA and TLA, is shown in Fig. \ref{FigESP}(b) and (c),
respectively. The plots in Fig. \ref{FigESP} (b) and (c) demonstrate that the
FMSR response to externally--applied driving is qualitatively similar to the
response of classical Mathieu and Duffing oscillators \cite{Landau1960a}.

Frequency mixing between simultaneously applied transverse and longitudinal
driving is demonstrated in two different configurations. In the first
configuration [see Fig. \ref{FigESP}(d)], the longitudinal driving angular
frequency, which is denoted by $\omega_{\mathrm{L}}$, is tuned close to
$2\omega_{\mathrm{T}}$, whereas $\omega_{\mathrm{L}}\ll\omega_{\mathrm{T}}$
for the second configuration [see Fig. \ref{FigESP} (e) and (f)]. For
practical applications, the first configuration is mainly used for
phase--sensitive amplification, whereas signal modulation can be implemented
using the second configuration \cite{Stancil_Spin}.

For the first configuration, for which $\omega_{\mathrm{L}}=2\left(
\omega_{\mathrm{T}}+\omega_{\mathrm{f}}\right)  $, the effect of the relative
phase $\phi_{\mathrm{T}}$ between the transverse and longitudinal driving
tones [see Eq. (\ref{H DLS})], is demonstrated using an intermodulation
measurement \cite{Mathai_67001}. The color coded plot in Fig. \ref{FigESP}(d)
displays the spectral peak intensity (measured using the RFSA) of the first
order frequency mixing between transverse and longitudinal driving tones,
which occurs at angular frequency $\omega_{\mathrm{L}}-\omega_{\mathrm{T}} =
\omega_{\mathrm{T}}+2\omega_{\mathrm{f}}$. For the second configuration, the
longitudinal driving frequency $\omega_{\mathrm{L}}/\left(  2\pi\right)  $ is
tuned to the value $0.5\operatorname{MHz}$. The measured RFSA trace shown in
Fig. \ref{FigESP} (e) contains a sequence of sidebands at angular frequencies
$\omega_{\mathrm{T}}+m\omega_{\mathrm{L}}$, where $m$ is an integer. For
comparison, a calculated spectral density, which is based on Eq. (D11) of Ref.
\cite{Buks_033807} (see also Ref. \cite{Shevchenko_1}), is shown in Fig.
\ref{FigESP} (f). Parameters' values are listed in the caption of Fig.
\ref{FigESP}.

\begin{figure}[ptb]
\begin{center}
\includegraphics[width=3.0in,keepaspectratio]{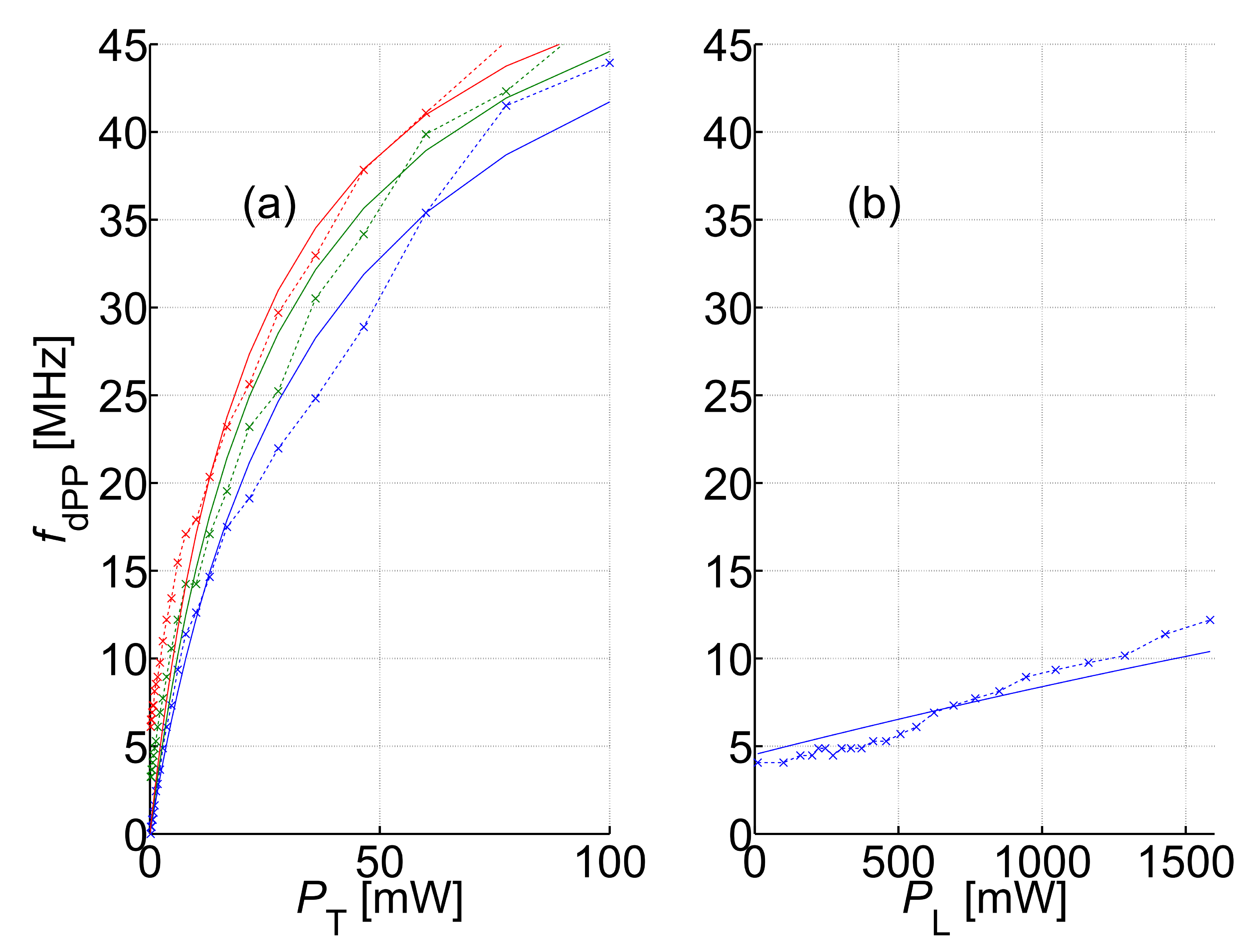}
\end{center}
\caption{{} Peak points. (a) The peak point
frequency shift $f_{\mathrm{dPP}}$ as a function of transverse driving power
$P_{\mathrm{T}}$ for longitudinal driving power $P_{\mathrm{L}}$ values of $0
\operatorname{mW}$ (blue), $320 \operatorname{mW}$ (green) and $560
\operatorname{mW}$ (red). (b) The dependency on longitudinal driving power
$P_{\mathrm{L}}$, for the case $P_{\mathrm{T}}=31.6\operatorname{mW}$. For
both plots, solid lines represent predictions derived using Eq. (\ref{dePP})
of SM section \ref{SM_RDM}. The dimensionless parameter $D$ in Eq.
(\ref{dePP}) is determined by measuring the transverse driving power and
frequency detuning at the lower bistability onset point (see SM section
\ref{SM_RDM}). A calibration yields the dimensionless parameters
$W=P_{\mathrm{T}}/\left(  37 \operatorname{mW}\right)  $ and $W_{\mathrm{A}%
}^{2}T_{2}^{2}=P_{\mathrm{L}}/\left(  980 \operatorname{mW}\right)  $.}%
\label{FigPP}%
\end{figure}

A quantitative comparison between data and predictions derived from the
disentanglement--based model is demonstrated by the plots shown in Fig.
\ref{FigPP}. The effect of driving on resonance line shape can be
characterized by the frequency shift of the peak (i.e. extremum) point, which
is denoted by $f_{\mathrm{dPP}}$. For the rapid disentanglement model,
$f_{\mathrm{dPP}}$ can be obtained from Eq. (\ref{dePP}) of SM section
\ref{SM_RDM}. A comparison between data and values derived from Eq.
(\ref{dePP}) is shown in Fig. \ref{FigPP}. Dependency on $P_{\mathrm{T}}$
(transverse driving power) and $P_{\mathrm{L}}$ (longitudinal driving power)
is shown in (a) and (b), respectively. Parameters' assumed values are listed
in the caption of Fig. \ref{FigPP}. The data--theory comparison demonstrates
that the disentanglement--based model is capable of qualitatively
accounting for the experimentally observed nonlinear response of
the spin system under study.

\textbf{Bosonization} -- While the linearity of standard QM excludes
multistability in finite quantum systems, some approximation methods can give
rise to nonlinear dynamics. The method of Bosonization introduces nonlinearity
that can give rise to bistability in the presence of magnetic anisotropy
\cite{Wang_224410,Buks_2400587}. However, it has remained unclear how this method, which
yields bistability that is otherwise excluded, can be justified \cite{Buks_2400587}.

The application of the Bosonization method to the under--study system of
parametrically driven spins is reviewed in SM section \ref{SM_MM}. For simplicity, it is assumed that FMSR magnetization is uniform (the validity
of this assumption is discussed in SM section \ref{SM_MM}). FMSR
damping is characterized by linear $\gamma=\gamma_{1}+\gamma_{2}$ and
nonlinear $\gamma_{3}$ rates, where $\gamma_{1}$ and $\gamma_{2}$\ represent,
respectively, the FMSR--TLA inductive coupling, and intrinsic FMSR loss.
Transverse driving is characterized by an amplitude $\omega_{\mathrm{T}1}$,
relative phase $\phi_{\mathrm{T}}$, and angular detuning frequency
$\Omega_{\mathrm{d}}$.

The Bosonization method yields in steady state a cubic polynomial equation for
the magnon number expectation value $E$, which is given by [see Eq.
(\ref{g_M}) in SM section \ref{SM_MM}]%
\begin{equation}
E=\frac{2\gamma_{1}\Omega_{1}}{\left(  \Omega_{\mathrm{d}}-\omega_{\mathrm{K}%
}E\right)  ^{2}+\left(  \gamma+\gamma_{3}E\right)  ^{2}}\;, \label{E MM SS}%
\end{equation}
where $\Omega_{1}=\omega_{\mathrm{T}1}g_{\mathrm{M}}$. The term $g_{\mathrm{M}%
}$ represents parametric gain, which periodically depends on the relative
phase $\phi_{\mathrm{T}}$ [see Eq. (\ref{g_M}) of SM section \ref{SM_MM}].
Note that a relation similar to Eq. (\ref{E MM SS}) is obtained from the
Mathieu model for the steady state of a parametrically driven classical
resonator \cite{Lifshitz_1}.

Stability analysis of the cubic polynomial equation (\ref{E MM SS}) has been
performed in Ref. \cite{Yurke_5054}. Bistability occurs provided that
$\left\vert \omega_{\mathrm{K}}\right\vert \geq\sqrt{3}\gamma_{3}$. In the
bistability region the cubic polynomial equation (\ref{E MM SS}) has three
real non--negative solutions (two of which represent locally stable steady
states). For $\left\vert \omega_{\mathrm{K}}\right\vert \geq\sqrt{3}\gamma
_{3}$, bistability is possible for driving amplitudes $\Omega_{1}$ bounded by
$\Omega_{1}\geq\Omega_{1\mathrm{c}}$, where $\Omega_{1\mathrm{c}%
}=E_{\mathrm{c}}^{3}\left(  \omega_{\mathrm{K}}^{2}+\gamma_{3}^{2}\right)
/\left(  2\gamma_{1}\right)  $ and $E_{\mathrm{c}}=\left(  2\gamma/\sqrt
{3}\right)  /\left(  \left\vert \omega_{\mathrm{K}}\right\vert -\sqrt{3}%
\gamma_{3}\right)  $.

Above parametric instability threshold, and in the absence of nonlinear
damping (i.e. for $\gamma_{3}=0$), the Bosonization--based model predicts that
the Bosonic mode amplitude exponentially increases as a function of time. A
finite steady state solution is obtained provided that $\gamma_{3}>0$.
Nonlinear damping in magnetically ordered dielectrics has been theoretically
studied in \cite{lvov_1562}. A pairing mechanism, which originates from a
fourth order interaction, suppresses pumping--induced magnon creation
\cite{Lvov_233}, and thus further bounds steady state amplitude.

The two competing theoretical models yield distinguishable predictions. For
example, the region where bistability occurs in the plane of driving amplitude
and driving frequency detuning is finite according to the rapid
disentanglement model (see Fig. \ref{FigRDz} in SM section \ref{SM_RDM}). In
contrast, the Bosonization--based model predicts an infinite region (both
driving amplitude $\Omega_{1}$ and detuning angular frequency $\Omega
_{\mathrm{d}}$ in the bistability region do not have upper bounds, see Fig.
\ref{FigME} in SM section \ref{SM_MM}).

Mapping measurements of the region of bistability are presented in SM section
\ref{SM_BS}. A comparison between data and theoretical predictions (see Fig.
\ref{FigSP} of SM section \ref{SM_BS}) reveals that the rapid disentanglement
model better aligns with data (in comparison with the Bosonization--based model).

\textbf{Discussion} -- The experimentally--observed effect of parallel pumping
on spins suggests that the underlying dynamics are nonlinear \cite{Lvov_233}.
The method of Bosonization can give rise to nonlinearity. However, the
justification of this method is arguably questionable
\cite{Katz_040404,Leppenen_2404_02134,Minganti_042118,Vicentini_013853,Landa_043601}%
, since it gives rise to multi-stability that is otherwise theoretically
excluded. An alternative model, which is based on the disentanglement master
equation (\ref{MME}), is found to be capable of qualitatively
accounting for both instability and bistability, which are
experimentally--observed in the spin system under study.

Generally, the modified master equation (\ref{MME}) can be constructed for any
physical system whose Hilbert space has finite dimensionality. The nonlinear
term added to the master equation (\ref{MME}) does not violate both norm
conservation and positivity of the density operator $\rho$ \cite{Buks_012439}.
For a multipartite system, partial disentanglement (i.e. disentanglement
between any pair of subsystems) can be introduced. In the absence of
entanglement (i.e. for product states), the nonlinear extension in the master
equation (\ref{MME}) does not vary any prediction of standard QM.
Disentanglement is invariant under any subsystem unitary transformation, and
it is applicable for both distinguishable and indistinguishable
particles \cite{Buks_630}. Moreover, the spontaneous disentanglement
hypothesis is falsifiable, since Eq. (\ref{MME}) yields predictions that are
distinguishable from what is derived from standard QM.

The spontaneous disentanglement hypothesis is arguably relevant to
the problem of quantum measurement
\cite{Penrose_4864,Bassi_471,Pearle_857,Ghirardi_470,Bassi_257,Bennett_170502,Kowalski_1,Fernengel_385701,Oppenheim_041040,Schrinski_133604}%
, which was first introduced in 1935 by Schr\"{o}dinger \cite{Schrodinger_807}%
. This problem was the main motivation for previously--proposed nonlinear
extensions to QM
\cite{Weinberg_61,Doebner_3764,Gisin_5677,Gisin_2259,Gisin_1,Kaplan_055002,Munoz_110503,Jacobs_279,Geller_2200156}%
. The nonlinear extension in the disentanglement master equation (\ref{MME})
makes the collapse postulate of QM redundant. Moreover, the spontaneous
disentanglement hypothesis is arguably related to phase transitions in quantum
systems \cite{Buks_012439} and to superconductivity \cite{Buks_630}.

\textbf{Summary} -- The current study compares predictions derived from two
competing theoretical models, with the measured FMSR response to
externally--applied transverse and longitudinal driving. Better agreement is
obtained from the disentanglement--based model (in comparison with the
Bosonization--based model). Further study is needed to determine whether the
spontaneous disentanglement hypothesis is consistent with experimental
observations obtained with other physical systems, and whether it is
internally consistent.

\textbf{Acknowledgments} -- The research was supported by the Technion
homeland security foundation.


\bibliographystyle{ieeepes}
\bibliography{acompat,Eyal_Bib}

\widetext \widetext \clearpage

\begin{center}
\textbf{{\large Supplementary materials: Parametric excitation of a
ferrimagnetic sphere resonator}}

{\small {Eyal Buks} }

{\small {Andrew and Erna Viterbi Department of Electrical Engineering,
Technion, Haifa 32000 Israel} }
\end{center}

\setcounter{equation}{0} \setcounter{figure}{0} \setcounter{table}{0}
\setcounter{section}{0} \setcounter{page}{1}
\makeatletter\renewcommand{\theequation}{S\arabic{equation}}
\renewcommand{\thefigure}{S\arabic{figure}}
\renewcommand{\thesection}{S\arabic{section}}
\renewcommand{\bibnumfmt}[1]{[S#1]} \renewcommand{\citenumfont}[1]{S#1}

Heisenberg equations of motion are derived in section \ref{SM_EOM}, and the
two--spin case $L=2$ is discussed in section \ref{SM_L2}. Sections
\ref{SM_RDM} and \ref{SM_MM} are devoted to the rapid disentanglement and the
Bosonization--based models, respectively. Predictions derived from these two
competing theoretical models are compared with experimental results in section
\ref{SM_BS}.

\section{Equations of motion}

\label{SM_EOM}

In this section, a unitary transformation into a rotating frame is
applied to simplify the equations of motion. The Hamiltonian
given by Eq. (\ref{H DLS}) in the main text yields Heisenberg equations of
motion given by%
\begin{align}
\frac{\mathrm{d}}{\mathrm{d}t}\left(
\begin{array}
[c]{c}%
S_{+}\\
S_{-}%
\end{array}
\right)   &  =\left(
\begin{array}
[c]{cc}%
-i\omega_{z} & 0\\
0 & i\omega_{z}%
\end{array}
\right)  \left(
\begin{array}
[c]{c}%
S_{+}\\
S_{-}%
\end{array}
\right) \nonumber\\
&  +S_{z}\left(
\begin{array}
[c]{cc}%
-\frac{i\omega_{\mathrm{K}}}{2} & -\frac{i\omega_{\mathrm{A}}}{2}\\
\frac{i\omega_{\mathrm{A}}}{2} & \frac{i\omega_{\mathrm{K}}}{2}%
\end{array}
\right)  \left(
\begin{array}
[c]{c}%
S_{+}\\
S_{-}%
\end{array}
\right)  +\left(
\begin{array}
[c]{cc}%
-\frac{i\omega_{\mathrm{K}}}{2} & -\frac{i\omega_{\mathrm{A}}}{2}\\
\frac{i\omega_{\mathrm{A}}}{2} & \frac{i\omega_{\mathrm{K}}}{2}%
\end{array}
\right)  \left(
\begin{array}
[c]{c}%
S_{+}\\
S_{-}%
\end{array}
\right)  S_{z}\nonumber\\
&  +\left(
\begin{array}
[c]{c}%
-iS_{z}\Omega_{\mathrm{T}1}^{\ast}e^{-i\omega_{\mathrm{T}}t}\\
iS_{z}\Omega_{\mathrm{T}1}^{{}}e^{i\omega_{\mathrm{T}}t}%
\end{array}
\right)  \;,\nonumber\\
&
\end{align}
and [note that $\left[  S_{+}S_{-}+S_{-}S_{+},S_{z}\right]  =0$, since
$S_{+}S_{-}+S_{-}S_{+}=2\left(  S_{x}^{2}+S_{y}^{2}\right)  $]%
\begin{equation}
\frac{\mathrm{d}S_{z}}{\mathrm{d}t}=-i\frac{\omega_{\mathrm{A}}\left(
S_{+}^{2}-S_{-}^{2}\right)  }{2}-i\frac{\left(  S_{+}\Omega_{\mathrm{T}1}^{{}%
}e^{i\omega_{\mathrm{T}}t}-S_{-}\Omega_{\mathrm{T}1}^{\ast}e^{-i\omega
_{\mathrm{T}}t}\right)  }{2}\;.
\end{equation}
The transformation%
\begin{equation}
\left(
\begin{array}
[c]{c}%
\mathcal{S}_{+}\\
\mathcal{S}_{-}%
\end{array}
\right)  =\left(
\begin{array}
[c]{cc}%
e^{i\omega_{\mathrm{T}}t} & 0\\
0 & e^{-i\omega_{\mathrm{T}}t}%
\end{array}
\right)  \left(
\begin{array}
[c]{cc}%
X & Y\\
Y & X
\end{array}
\right)  \left(
\begin{array}
[c]{c}%
S_{+}\\
S_{-}%
\end{array}
\right)  \;, \label{S pm T DLS}%
\end{equation}
where $X=\left(  1/2\right)  \left(  \sqrt{1+\omega_{\mathrm{A}}%
/\omega_{\mathrm{K}}}+\sqrt{1-\omega_{\mathrm{A}}/\omega_{\mathrm{K}}}\right)
$ and where $Y=\left(  1/2\right)  \left(  \sqrt{1+\omega_{\mathrm{A}}%
/\omega_{\mathrm{K}}}-\sqrt{1-\omega_{\mathrm{A}}/\omega_{\mathrm{K}}}\right)
$, yields [note that $X^{2}+Y^{2}=1$, $2XY=\omega_{\mathrm{A}}/\omega
_{\mathrm{K}}$, and $X^{2}-Y^{2}=\sqrt{1-\left(  \omega_{\mathrm{A}}%
/\omega_{\mathrm{K}}\right)  ^{2}}=1/\varrho$, where $\varrho=1/\sqrt
{1-\left(  \omega_{\mathrm{A}}/\omega_{\mathrm{K}}\right)  ^{2}}$]%
\begin{align}
\frac{\mathrm{d}}{\mathrm{d}t}\left(
\begin{array}
[c]{c}%
\mathcal{S}_{+}\\
\mathcal{S}_{-}%
\end{array}
\right)   &  =i\left(
\begin{array}
[c]{cc}%
\omega_{\mathrm{T}}-\varrho\omega_{z} & \varrho\omega_{z}\frac{\omega
_{\mathrm{A}}}{\omega_{\mathrm{K}}}e^{2i\omega_{\mathrm{T}}t}\\
-\varrho\omega_{z}\frac{\omega_{\mathrm{A}}}{\omega_{\mathrm{K}}}%
e^{-2i\omega_{\mathrm{T}}t} & -\omega_{\mathrm{T}}+\varrho\omega_{z}%
\end{array}
\right)  \left(
\begin{array}
[c]{c}%
\mathcal{S}_{+}\\
\mathcal{S}_{-}%
\end{array}
\right) \nonumber\\
&  +\frac{i\omega_{\mathrm{K}}}{2\varrho}\left(
\begin{array}
[c]{c}%
-\mathcal{S}_{+}S_{z}-S_{z}\mathcal{S}_{+}\\
\mathcal{S}_{-}S_{z}+S_{z}\mathcal{S}_{-}%
\end{array}
\right) \nonumber\\
&  +i\left(
\begin{array}
[c]{c}%
-X\Omega_{\mathrm{T}1}^{\ast}+Y\Omega_{\mathrm{T}1}^{{}}e^{2i\omega
_{\mathrm{T}}t}\\
X\Omega_{\mathrm{T}1}^{{}}-Y\Omega_{\mathrm{T}1}^{\ast}e^{-2i\omega
_{\mathrm{T}}t}%
\end{array}
\right)  S_{z}\;,\nonumber\\
&  \label{d/dt S pm DLS}%
\end{align}
and%
\begin{align}
\frac{\mathrm{d}S_{z}}{\mathrm{d}t}  &  =\frac{i\varrho\omega_{\mathrm{A}%
}\left(  e^{2i\omega_{\mathrm{T}}t}\mathcal{S}_{-}^{2}-e^{-2i\omega
_{\mathrm{T}}t}\mathcal{S}_{+}^{2}\right)  }{2}\nonumber\\
&  -i\varrho\frac{X\left(  \Omega_{\mathrm{T}1}^{{}}\mathcal{S}_{+}%
-\Omega_{\mathrm{T}1}^{\ast}\mathcal{S}_{-}\right)  +Y\left(  \Omega
_{\mathrm{T}1}^{\ast}e^{-2i\omega_{\mathrm{T}}t}\mathcal{S}_{+}-\Omega
_{\mathrm{T}1}^{{}}e^{2i\omega_{\mathrm{T}}t}\mathcal{S}_{-}\right)  }%
{2}\;.\nonumber\\
&  \label{d/dt S z DLS}%
\end{align}

In the rotating wave approximation (RWA) Eqs. (\ref{d/dt S pm DLS}) and
(\ref{d/dt S z DLS}) become [recall that $\omega_{\mathrm{T}}=\omega
_{0}+\omega_{\mathrm{d}}$ and $\omega_{z}=\omega_{0}+\Omega_{\mathrm{L}1}^{{}%
}\cos\left(  2\left(  \omega_{\mathrm{T}}+\omega_{\mathrm{f}}\right)
t\right)  $, and note that it is assumed that $\left\vert \omega_{\mathrm{d}%
}\right\vert \ll\omega_{0}$ and $\left\vert \omega_{\mathrm{f}}\right\vert
\ll\omega_{0}$]%
\begin{align}
\frac{\mathrm{d}}{\mathrm{d}t}\left(
\begin{array}
[c]{c}%
\mathcal{S}_{+}\\
\mathcal{S}_{-}%
\end{array}
\right)   &  =i\left(
\begin{array}
[c]{cc}%
W_{\mathrm{d}} & W_{\mathrm{A}}e^{-2i\omega_{\mathrm{f}}t}\\
-W_{\mathrm{A}}e^{2i\omega_{\mathrm{f}}t} & -W_{\mathrm{d}}%
\end{array}
\right)  \left(
\begin{array}
[c]{c}%
\mathcal{S}_{+}\\
\mathcal{S}_{-}%
\end{array}
\right) \nonumber\\
&  +\frac{iW_{\mathrm{K}}}{2}\left(
\begin{array}
[c]{c}%
-\mathcal{S}_{+}S_{z}-S_{z}\mathcal{S}_{+}\\
\mathcal{S}_{-}S_{z}+S_{z}\mathcal{S}_{-}%
\end{array}
\right) \nonumber\\
&  +i\left(
\begin{array}
[c]{c}%
-W_{\mathrm{T}1}^{\ast}S_{z}\\
W_{\mathrm{T}1}^{{}}S_{z}%
\end{array}
\right)  \;,\nonumber\\
&  \label{d Spm / dt DLS}%
\end{align}
and%
\begin{equation}
\frac{\mathrm{d}S_{z}}{\mathrm{d}t}=\frac{i\varrho\left(  W_{\mathrm{T}%
1}^{\ast}\mathcal{S}_{-}-W_{\mathrm{T}1}^{{}}\mathcal{S}_{+}\right)  }{2}\;.
\label{d Sz / dt DLS}%
\end{equation}
where $W_{\mathrm{d}}=\omega_{\mathrm{T}}-\varrho\omega_{0}$, $W_{\mathrm{A}%
}=\left(  \varrho/2\right)  \left(  \omega_{\mathrm{A}}/\omega_{\mathrm{K}%
}\right)  \Omega_{\mathrm{L}1}^{{}}$, $W_{\mathrm{K}}=\omega_{\mathrm{K}%
}/\varrho$ and $W_{\mathrm{T}1}=\Omega_{\mathrm{T}1}^{{}}X$. Note
that for $\omega_{\mathrm{f}}=0$ (i.e. vanishing longitudinal driving angular
detuning frequency), the equation of motion (\ref{d Spm / dt DLS}) becomes
time independent.

\section{The case $L=2$}

\label{SM_L2}

For sufficiently small number of spins $L$, some analytical
results can be derived. The smallest value, for which disentanglement is
relevant, is $L=2$ (i.e. two spins). For that case, the matrix
representation of the Hamiltonian $\mathcal{H}$ [see Eq. (\ref{H DLS}) in the
main text] is given by (recall that $\left[  S_{i},S_{j}\right]
=2i\epsilon_{ijk}S_{k}$, $\left[  S_{z},S_{\pm}\right]  =\pm2S_{\pm}$ and
$\left[  S_{+},S_{-}\right]  =4S_{z}$, where $S_{\pm}=S_{x}\pm iS_{y}$)%
\begin{equation}
\frac{\mathcal{H}}{\hbar}\dot{=}\left(
\begin{array}
[c]{cccc}%
-\omega_{z}+\omega_{\mathrm{K}} & \omega_{\mathrm{t}}^{{}} & \omega
_{\mathrm{t}}^{{}} & \omega_{\mathrm{A}}\\
\omega_{\mathrm{t}}^{\ast} & \omega_{\mathrm{K}} & \omega_{\mathrm{K}} &
\omega_{\mathrm{t}}^{{}}\\
\omega_{\mathrm{t}}^{\ast} & \omega_{\mathrm{K}} & \omega_{\mathrm{K}} &
\omega_{\mathrm{t}}^{{}}\\
\omega_{\mathrm{A}} & \omega_{\mathrm{t}}^{\ast} & \omega_{\mathrm{t}}^{\ast}
& \omega_{z}+\omega_{\mathrm{K}}%
\end{array}
\right)  \;, \label{H 4X4}%
\end{equation}
where $\omega_{\mathrm{t}}=\left(  \Omega_{\mathrm{T}1}^{{}}/2\right)
e^{i\omega_{\mathrm{T}}t}$. For the case $\omega_{\mathrm{K}}=0$,
$\omega_{\mathrm{t}}=0$ and $\omega_{\mathrm{T}}=\omega_{0}$, the
non--vanishing entries of the Hamiltonian $\mathcal{H}$ (\ref{H 4X4}) can be
represented by a $2\times2$ block given by [recall that $\omega_{z}=\omega
_{0}+\Omega_{\mathrm{L}1}^{{}}\cos\left(  2\omega_{0}t\right)  $]%
\begin{equation}
\frac{\mathcal{H}}{\hbar}\dot{=}-\omega_{0}\left(
\begin{array}
[c]{cc}%
1 & \tan q\\
\tan q & -1
\end{array}
\right)  -\Omega_{\mathrm{L}1}^{{}}\cos\left(  2\omega_{0}t\right)  \left(
\begin{array}
[c]{cc}%
1 & 0\\
0 & -1
\end{array}
\right)  \;,
\end{equation}
where $\tan q=-\omega_{\mathrm{A}}/\omega_{0}$.

Diagonalization of the static part is performed by the transformation%
\begin{align}
\frac{\mathcal{H}^{\prime}}{\hbar}  &  =u^{-1}\frac{\mathcal{H}}{\hbar}u^{{}%
}\dot{=}-\frac{\omega_{0}}{\cos q}\left(
\begin{array}
[c]{cc}%
1 & 0\\
0 & -1
\end{array}
\right) \nonumber\\
&  -\Omega_{\mathrm{L}1}^{{}}\cos\left(  2\omega_{0}t\right)  \left(
\begin{array}
[c]{cc}%
\cos q & -\sin q\\
-\sin q & -\cos q
\end{array}
\right)  \;,\nonumber\\
&
\end{align}
where%
\begin{equation}
u\dot{=}\left(
\begin{array}
[c]{cc}%
\cos\frac{q}{2} & -\sin\frac{q}{2}\\
\sin\frac{q}{2} & \cos\frac{q}{2}%
\end{array}
\right)  \;.
\end{equation}
A rotating frame transformation yields%
\begin{align}
\frac{\mathcal{H}_{{}}^{\prime\prime}}{\hbar}  &  =-iu_{0}^{\dag}%
\frac{\mathrm{d}u_{0}^{{}}}{\mathrm{d}t}+u_{0}^{\dag}\frac{\mathcal{H}_{{}%
}^{\prime}}{\hbar}u_{0}^{{}}\nonumber\\
&  \dot{=}\omega_{0}\frac{\cos q-1}{\cos q}\left(
\begin{array}
[c]{cc}%
1 & 0\\
0 & -1
\end{array}
\right) \nonumber\\
&  -\Omega_{\mathrm{L}1}^{{}}\cos\left(  2\omega_{0}t\right)  \left(
\begin{array}
[c]{cc}%
\cos q & -e^{-2i\omega_{0}t}\sin q\\
-e^{2i\omega_{0}t}\sin q & -\cos q
\end{array}
\right)  \;,\nonumber\\
&
\end{align}
where%
\begin{equation}
u_{0}\dot{=}\left(
\begin{array}
[c]{cc}%
e^{i\omega_{0}t} & 0\\
0 & e^{-i\omega_{0}t}%
\end{array}
\right)  \;.
\end{equation}
Applying the RWA leads to the Hamiltonian $\mathcal{H}_{\mathrm{RWA}}%
^{\prime\prime}$, which has a matrix representation given by%
\begin{equation}
\frac{\mathcal{H}_{\mathrm{RWA}}^{\prime\prime}}{\hbar}\dot{=}\omega_{0}%
\frac{\cos q-1}{\cos q}\left(
\begin{array}
[c]{cc}%
1 & 0\\
0 & -1
\end{array}
\right)  +\frac{\Omega_{\mathrm{L}1}^{{}}\sin q}{2}\left(
\begin{array}
[c]{cc}%
0 & 1\\
1 & 0
\end{array}
\right)  \;.
\end{equation}
The transformation inverse to $u$ yields%
\begin{align}
\frac{\mathcal{H}^{\prime\prime\prime}}{\hbar}  &  =u^{{}}\frac{\mathcal{H}%
_{\mathrm{RWA}}^{\prime\prime}}{\hbar}u^{-1}\nonumber\\
&  \dot{=}\omega_{0}\left(  \cos q-1\right)  \left(
\begin{array}
[c]{cc}%
1 & \tan q\\
\tan q & -1
\end{array}
\right) \nonumber\\
&  +\frac{\Omega_{\mathrm{L}1}\sin q}{2}\left(
\begin{array}
[c]{cc}%
-\sin q & \cos q\\
\cos q & \sin q
\end{array}
\right)  \;.\nonumber\\
&
\end{align}
For $\Omega_{\mathrm{L}1}\ll\omega_{0}$ and $\omega_{\mathrm{A}}\ll\omega_{0}%
$, ${\mathcal{H}^{\prime\prime\prime}}$ has a $4\times4$ matrix representation
given by [recall that $q=\tan^{-1}\left(  -\omega_{\mathrm{A}}/\omega
_{0}\right)  =-\omega_{\mathrm{A}}/\omega_{0}+O\left(  \omega_{\mathrm{A}}%
^{3}\right)  $]%
\begin{equation}
\frac{\mathcal{H}^{\prime\prime\prime}}{\hbar}\dot{=}-\frac{\omega
_{\mathrm{A}}}{2\omega_{0}}\left(
\begin{array}
[c]{cccc}%
\omega_{\mathrm{A}} & 0 & 0 & \Omega_{\mathrm{L}1}\\
0 & 0 & 0 & 0\\
0 & 0 & 0 & 0\\
\Omega_{\mathrm{L}1} & 0 & 0 & -\omega_{\mathrm{A}}%
\end{array}
\right)  \;.
\end{equation}

\section{Rapid disentanglement approximation}

\label{SM_RDM}

The spontaneous disentanglement hypothesis is formulated using a
nonlinear master equation given by Eq. (\ref{MME}) in the main text. For large
systems, this nonlinear master equation is commonly intractable. In the
current section, the rapid disentanglement approximation, which greatly
simplifies the dynamics, is explored.

The term $\mathcal{S}_{\pm}S_{z}+S_{z}\mathcal{S}_{\pm}$ in Eq.
(\ref{d Spm / dt DLS}) can be expressed as $\mathcal{S}_{\pm}S_{z}%
+S_{z}\mathcal{S}_{\pm}=\sum_{l^{\prime},l^{\prime\prime}=1}^{L}\left(
\mathcal{S}_{l^{\prime},\pm}S_{l^{\prime\prime},z}+S_{z,l^{\prime}}%
\mathcal{S}_{l^{\prime},\pm}\right)  $. The approximation $\left\langle
\mathcal{S}_{l^{\prime},\pm}S_{l^{\prime\prime},z}+S_{z,l^{\prime}}%
\mathcal{S}_{l^{\prime},\pm}\right\rangle \simeq\left\langle \mathcal{S}%
_{l^{\prime},\pm}\right\rangle \left\langle S_{l^{\prime\prime},z}%
\right\rangle +\left\langle S_{z,l^{\prime}}\right\rangle \left\langle
\mathcal{S}_{l^{\prime\prime},\pm}\right\rangle \simeq2L^{-2}\left\langle
\mathcal{S}_{\pm}\right\rangle \left\langle S_{z}\right\rangle $ can be
implemented provided that the rate of disentanglement $\gamma_{\mathrm{D}}%
$\ is sufficiently large. By implementing this approximation, and by averaging
and adding damping terms to Eqs. (\ref{d Spm / dt DLS}) and
(\ref{d Sz / dt DLS}), one obtains%
\begin{align}
&  \frac{\mathrm{d}}{\mathrm{d}t}\left(
\begin{array}
[c]{c}%
P_{+}\\
P_{-}%
\end{array}
\right) \nonumber\\
&  =\left(
\begin{array}
[c]{cc}%
i\left(  W_{\mathrm{d}}-W_{\mathrm{K}}P_{z}\right)  -\frac{1}{T_{2}} &
iW_{\mathrm{A}}e^{-2i\omega_{\mathrm{f}}t}\\
-iW_{\mathrm{A}}e^{2i\omega_{\mathrm{f}}t} & -i\left(  W_{\mathrm{d}%
}-W_{\mathrm{K}}P_{z}\right)  -\frac{1}{T_{2}}%
\end{array}
\right)  \left(
\begin{array}
[c]{c}%
P_{+}\\
P_{-}%
\end{array}
\right) \nonumber\\
&  +iP_{z}\left(
\begin{array}
[c]{c}%
-W_{\mathrm{T}1}^{\ast}\\
W_{\mathrm{T}1}^{{}}%
\end{array}
\right)  \;,\nonumber\\
&  \label{d/dt P pm DLS}%
\end{align}
and%
\begin{equation}
\frac{\mathrm{d}P_{z}}{\mathrm{d}t}=\frac{i\varrho\left(  W_{\mathrm{T}%
1}^{\ast}P_{-}-W_{\mathrm{T}1}^{{}}P_{+}\right)  }{2}-\frac{P_{z}-P_{z0}%
}{T_{1}}\;, \label{d/dt Pz DLS}%
\end{equation}
where $P_{\pm}=\left\langle \mathcal{S}_{\pm}\right\rangle $ and
$P_{z}=\left\langle S_{z}\right\rangle $. Note that Eqs. (\ref{d/dt P pm DLS})
and (\ref{d/dt Pz DLS}) are invariant under the transformation $\left(
\phi_{\mathrm{T}},P_{+},P_{-}\right)  \rightarrow\left(  \phi_{\mathrm{T}}%
+\pi,-P_{+},-P_{-}\right)  $ (recall that $W_{\mathrm{T}1}=X\left\vert
\Omega_{\mathrm{T}1}\right\vert e^{i\phi_{\mathrm{T}}}$). Note
that the same invariance occurs in the classical realm, which is described by
the Mathieu model.

For the case $\omega_{\mathrm{f}}=0$ (no parametric detuning) and
$W_{\mathrm{T}1}=0$ (no transverse driving), Eq. (\ref{d/dt P pm DLS}) yields
$\mathrm{d}\mathcal{N}/\mathrm{d}t=2\left(  W_{\mathrm{A}}\sin\left(
2\phi_{\mathrm{P}}\right)  -1/T_{2}\right)  \mathcal{N}$, where $\mathcal{N}%
\equiv P_{+}P_{-}=P_{x}^{2}+P_{y}^{2}$, $P_{\pm}=\sqrt{\mathcal{N}}e^{\pm
i\phi_{\mathrm{P}}}$, and $\phi_{\mathrm{P}}$ is real. Phase--dependent rate of magnon creation by parallel pumping is represented by
the term $W_{\mathrm{A}}\sin\left(  2\phi_{\mathrm{P}}\right)  $ [recall that
$W_{\mathrm{A}}=\left(  \varrho/2\right)  \left(  \omega_{\mathrm{A}}%
/\omega_{\mathrm{K}}\right)  \Omega_{\mathrm{L}1}^{{}}$].

\subsection{The case $\omega_{\mathrm{f}}=0$}

For the case $\omega_{\mathrm{f}}=0$ (i.e. no parametric
detuning), a unitary transformation given by%
\begin{equation}
\left(
\begin{array}
[c]{c}%
P_{x}\\
P_{y}%
\end{array}
\right)  =\left(
\begin{array}
[c]{cc}%
1 & i\\
1 & -i
\end{array}
\right)  ^{-1}\left(
\begin{array}
[c]{c}%
P_{+}\\
P_{-}%
\end{array}
\right)  \;,
\end{equation}
yields real equations of motion given by [see Eqs. (\ref{d/dt P pm DLS}) and
(\ref{d/dt Pz DLS})]%
\begin{equation}
\frac{\mathrm{d}\mathbf{P}}{\mathrm{d}t}=\left(
\begin{array}
[c]{ccc}%
-\frac{1}{T_{2}} & -w_{\mathrm{d}}+W_{\mathrm{A}} & -W_{\mathrm{T}1\mathrm{I}%
}^{{}}\\
w_{\mathrm{d}}+W_{\mathrm{A}} & -\frac{1}{T_{2}} & -W_{\mathrm{T}1\mathrm{R}%
}^{{}}\\
\varrho W_{\mathrm{T}1\mathrm{I}}^{{}} & \varrho W_{\mathrm{T}1\mathrm{R}}%
^{{}} & -\frac{1}{T_{1}}%
\end{array}
\right)  \mathbf{P}+\left(
\begin{array}
[c]{c}%
0\\
0\\
\frac{P_{z0}}{T_{1}}%
\end{array}
\right)  \;,
\end{equation}
where $\mathbf{P}=\left(  P_{x},P_{y},P_{z}\right)  ^{\mathrm{T}}$, and where%
\begin{align}
W_{\mathrm{T}1\mathrm{R}}^{{}}  &  =\frac{W_{\mathrm{T}1}^{{}}+W_{\mathrm{T}%
1}^{\ast}}{2}\;,\\
W_{\mathrm{T}1\mathrm{I}}^{{}}  &  =\frac{W_{\mathrm{T}1}^{{}}-W_{\mathrm{T}%
1}^{\ast}}{2i}\;.
\end{align}
Alternatively%
\begin{equation}
\frac{\mathrm{d}\mathbf{P}}{\mathrm{d}t}=\mathbf{W}_{\mathrm{R}}%
\times\mathbf{P}+M_{\mathrm{d}}\mathbf{P}+\left(
\begin{array}
[c]{c}%
0\\
0\\
\frac{P_{z0}}{T_{1}}%
\end{array}
\right)  \;,
\end{equation}
where the vector $\mathbf{W}_{\mathrm{R}}$, which is given by $\mathbf{W}%
_{\mathrm{R}}=\left(  W_{\mathrm{T}1\mathrm{R}}^{{}},-W_{\mathrm{T}%
1\mathrm{I}}^{{}},w_{\mathrm{d}}\right)  ^{\mathrm{T}}$, represents a rotation
axis, and where the matrix $M_{\mathrm{d}}$ is given by (note that
$W_{\mathrm{A}}=0$ and $\varrho-1=0$ when $\omega_{\mathrm{A}}=0$)%
\begin{equation}
M_{\mathrm{d}}=\left(
\begin{array}
[c]{ccc}%
-\frac{1}{T_{2}} & W_{\mathrm{A}} & 0\\
W_{\mathrm{A}} & -\frac{1}{T_{2}} & 0\\
\left(  \varrho-1\right)  W_{\mathrm{T}1\mathrm{I}}^{{}} & \left(
\varrho-1\right)  W_{\mathrm{T}1\mathrm{R}}^{{}} & -\frac{1}{T_{1}}%
\end{array}
\right)  \;.
\end{equation}
The eigenvalues of the matrix $M_{\mathrm{d}}$, which represent effective
values of spin damping rates, are $-\left(  1+W_{\mathrm{A}}T_{2}\right)
/T_{2}$, $-\left(  1-W_{\mathrm{A}}T_{2}\right)  /T_{2}$ and $-1/T_{1}$.

\subsection{Bistability for the case $\omega_{\mathrm{f}}=0$}

Steady states, which are time--independent solutions of the
equations of motion, are derived in this section. For
$\omega_{\mathrm{f}}=0$ (i.e. vanishing longitudinal driving angular detuning
frequency), Eq. (\ref{d/dt P pm DLS}) can be expressed as%
\begin{equation}
\frac{\mathrm{d}}{\mathrm{d}t}\left(
\begin{array}
[c]{c}%
P_{+}\\
P_{-}%
\end{array}
\right)  =\frac{1}{T_{2}}M_{\mathrm{T}}\left(
\begin{array}
[c]{c}%
P_{+}\\
P_{-}%
\end{array}
\right)  +izP_{z0}\left(
\begin{array}
[c]{c}%
-W_{\mathrm{T}1}^{\ast}\\
W_{\mathrm{T}1}^{{}}%
\end{array}
\right)  \;, \label{d/dt P pm M_T DLS}%
\end{equation}
where the matrix $M_{\mathrm{T}}$ is given by%
\begin{equation}
M_{\mathrm{T}}=\left(
\begin{array}
[c]{cc}%
i\left(  \delta-4\sqrt{D}z\right)  -1 & i\sqrt{1-\alpha}\\
-i\sqrt{1-\alpha} & -i\left(  \delta-4\sqrt{D}z\right)  -1
\end{array}
\right)  \;, \label{M_T DLS}%
\end{equation}
$\delta=W_{\mathrm{d}}T_{2}$, $D=\left(  W_{\mathrm{K}}T_{2}P_{z0}/4\right)
^{2}$, $z=P_{z}/P_{z0}$ and $\alpha=1-W_{\mathrm{A}}^{2}T_{2}^{2}$. Steady
state of Eq. (\ref{d/dt P pm M_T DLS}) is given by%
\begin{equation}
\left(
\begin{array}
[c]{c}%
P_{+}\\
P_{-}%
\end{array}
\right)  =izP_{z0}T_{2}M_{\mathrm{T}}^{-1}\left(
\begin{array}
[c]{c}%
W_{\mathrm{T}1}^{\ast}\\
-W_{\mathrm{T}1}^{{}}%
\end{array}
\right)  \;, \label{v P pm DLS}%
\end{equation}
and thus (recall that $W_{\mathrm{T}1}=\left\vert W_{\mathrm{T}1}\right\vert
e^{i\phi_{\mathrm{T}}}$)%
\begin{align}
W_{\mathrm{T}1}^{\ast}P_{-}-W_{\mathrm{T}1}^{{}}P_{+}  &  =2izP_{z0}%
T_{2}\left\vert W_{\mathrm{T}1}^{{}}\right\vert ^{2}\frac{1+\sqrt{1-\alpha
}\sin\left(  2\phi_{\mathrm{T}}\right)  }{\alpha+\left(  \delta-4\sqrt
{D}z\right)  ^{2}}\;,\nonumber\\
&
\end{align}
hence in steady state [see Eq. (\ref{d/dt Pz DLS})]%
\begin{equation}
z=\frac{1}{1+\frac{2W}{\alpha+\left(  \delta-4\sqrt{D}z\right)  ^{2}}}\;,
\label{z= DLS}%
\end{equation}
where $W=\left(  1/2\right)  \varrho\left\vert W_{\mathrm{T}1}^{{}}\right\vert
^{2}T_{1}T_{2}\left[  1+\sqrt{1-\alpha}\sin\left(  2\phi_{\mathrm{T}}\right)
\right]  $. The cubic polynomial equation for $z$ (\ref{z= DLS}) can be
expressed as $0=F\left(  z,\delta\right)  $, where%
\begin{equation}
F\left(  z,\delta\right)  =z\left(  \alpha+\left(  \delta-4\sqrt{D}z\right)
^{2}+2W\right)  -\alpha-\left(  \delta-4\sqrt{D}z\right)  ^{2}\;.
\label{F(z,de)}%
\end{equation}

\textbf{Peak points} -- At peak points, for which $0=\mathrm{d}z/\mathrm{d}%
\delta=-\left(  \partial F/\partial\delta\right)  /\left(  \partial F/\partial
z\right)  $, where $\partial F/\partial\delta=2\left(  z-1\right)  \left(
\delta-4\sqrt{D}z\right)  $, the condition $0=F$ yields $z=1/\left(  1+\left(
2W/\alpha\right)  \right)  $ and%
\begin{equation}
\delta=\frac{4\sqrt{D}}{1+\frac{2W}{\alpha}}\;. \label{dePP}%
\end{equation}

\textbf{Bistability onset points} -- At a bistability onset point the
following three conditions hold%
\begin{equation}
0=\frac{\mathrm{d}\delta}{\mathrm{d}z}=-\frac{F_{z}}{F_{\delta}}\;, \label{C1}%
\end{equation}%
\begin{equation}
0=\frac{\mathrm{d}^{2}\delta}{\mathrm{d}z^{2}}=-\frac{F_{\delta}^{2}%
F_{zz}-2F_{z}F_{\delta}F_{z\delta}+F_{z}^{2}F_{\delta\delta}}{F_{\delta}^{3}%
}\;, \label{C2}%
\end{equation}
and%
\begin{equation}
0=F\left(  z,\delta\right)  \;, \label{C3}%
\end{equation}
where $F$ with an added subscript denotes a partial derivative, e.g.
$F_{z}=\partial F/\partial z$. Solution for $z$ of conditions (\ref{C1}),
(\ref{C2}) and (\ref{C3}) are given by $z_{1}=Z\left(  q\right)  $, and
$z_{\pm}=Z\left(  qe^{\pm2\pi i/3}\right)  $, where the function $Z\left(
x\right)  $ is defined by $Z\left(  x\right)  =\left(  x+1/x+3\right)  /4$,
and where $q=\exp\left(  i\left(  2/3\right)  \cos^{-1}\sqrt{\alpha/D}\right)
$. For any given solution for $z$, the corresponding variables $\delta$ and
$W$ are given by $\delta=2\sqrt{D}(3z-1)$ and $W=6D\left(  1-z\right)
^{2}-\alpha/2$. For the case $1\leq\alpha/D$ bistability is excluded. For the
case $\alpha/D=1$, the following holds $z_{\pm}=1/2$, $\delta=\sqrt{\alpha}$,
and $W=\alpha$. For negative $\alpha$, only a single bistability onset point
exists, and for this case bistability occurs only \textit{below} the
corresponding critical value of $W$.

\begin{figure}[ptb]
\begin{center}
\includegraphics[width=3.0in,keepaspectratio]{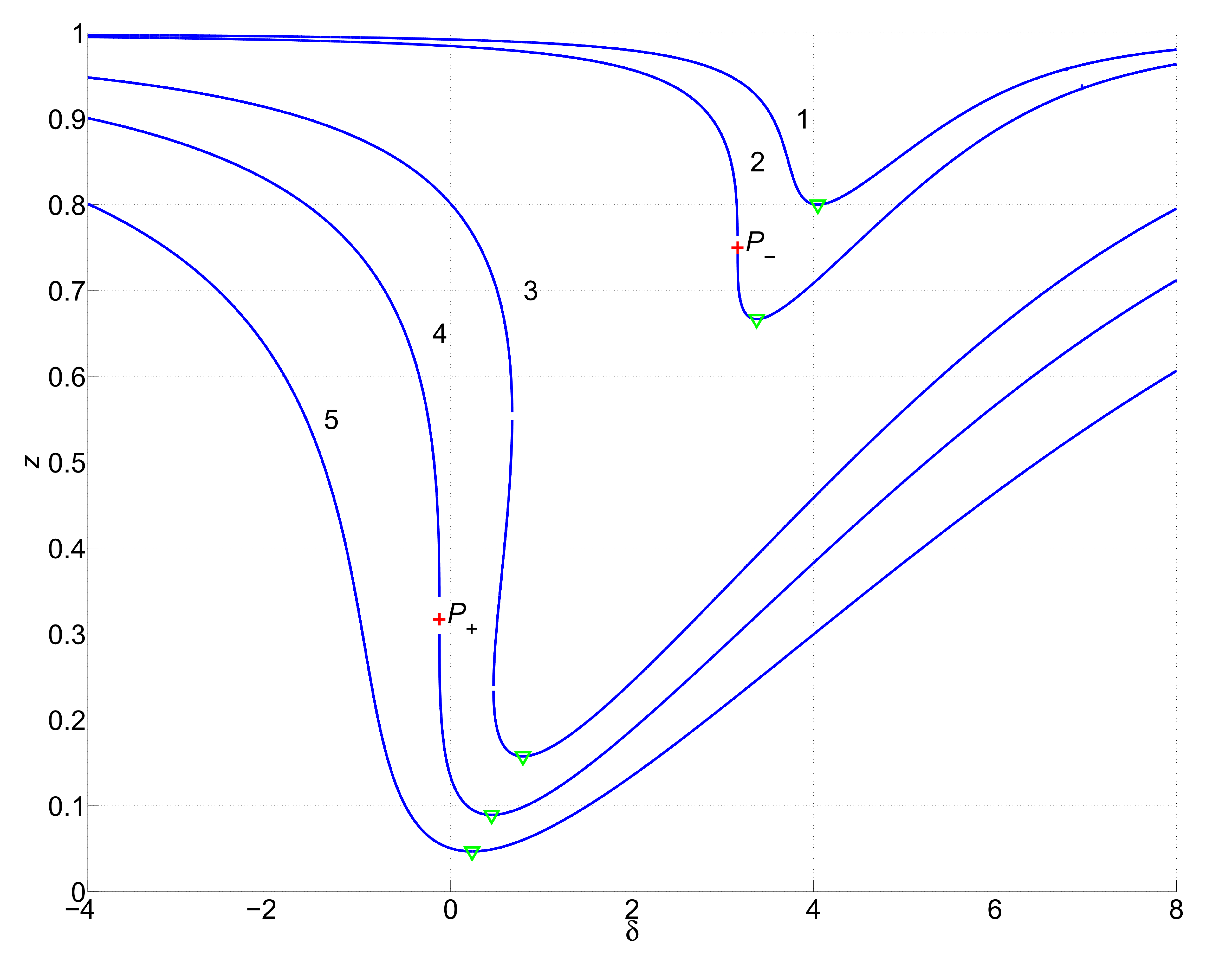}
\end{center}
\caption{{}Dimensionless polarization$\ z$ as a function of dimensionless
detuning $\delta$ for the rapid disentanglement model. Calculation of steady
state is based on Eq. (\ref{z= DLS}). Assumed parameters's values are
$\alpha=0.8$ and $D=2\alpha$. The five curves, which are labelled by the
numbers 1, 2, 3, 4, and 5, are calculated for five different values of the
dimensionless driving amplitude $W$ respectively given by $W_{-}/2$, $W_{-}$,
$\left(  W_{-}+W_{+}\right)  /2$, $W_{+}$ and $2W_{+}$, where $W_{\pm}$ is the
value of $W$ corresponding to the bistability onset point having value of $z$
given by $z_{\pm}$. Peak points are labeled by green triangles [see Eq.
(\ref{dePP})].}%
\label{FigRDz}%
\end{figure}

The plot in Fig. \ref{FigRDz}, which is based on the cubic polynomial equation
(\ref{F(z,de)}), displays steady state values of the dimensionless
polarization$\ z$ as a function of the dimensionless detuning $\delta$, for
five different values of the dimensionless driving amplitude $W$. The symbol
$P_{\pm}$ labels the bistability onset point having dimensionless
polarization$\ z_{\pm}$ and dimensionless detuning $\delta_{\pm}$. Peak points
are labeled by green triangles [see Eq. (\ref{dePP})].

\subsection{Parametric gain for the case $\omega_{\mathrm{f}}=0$}

The eigenvalues of the matrix $M_{\mathrm{T}}$ (\ref{M_T DLS}) are $-1\pm
\sqrt{1-\alpha-\left(  \delta-4\sqrt{D}z\right)  ^{2}}$. This matrix can be
decomposed as%
\begin{align}
M_{\mathrm{T}}  &  =\left(
\begin{array}
[c]{cc}%
\mu_{1}^{{}} & \mu_{2}^{{}}\\
\mu_{2}^{\ast} & \mu_{1}^{\ast}%
\end{array}
\right) \nonumber\\
&  =\left(
\begin{array}
[c]{cc}%
\left\vert \mu_{1}\right\vert e^{i\phi_{1}} & \left\vert \mu_{2}\right\vert
e^{i\phi_{2}}\\
\left\vert \mu_{2}\right\vert e^{-i\phi_{2}} & \left\vert \mu_{1}\right\vert
e^{-i\phi_{1}}%
\end{array}
\right) \nonumber\\
&  =R^{{}}\left(  \frac{\phi_{1}+\phi_{2}}{2}\right)  \left(
\begin{array}
[c]{cc}%
\left\vert \mu_{1}\right\vert +\left\vert \mu_{2}\right\vert  & 0\\
0 & \left\vert \mu_{1}\right\vert -\left\vert \mu_{2}\right\vert
\end{array}
\right)  R^{\dag}\left(  \frac{\phi_{2}-\phi_{1}}{2}\right)  \;,\nonumber\\
&  \label{M_T mu12}%
\end{align}
where $\mu_{1}=i\left(  \delta-4\sqrt{D}z\right)  -1=\left\vert \mu
_{1}\right\vert e^{i\phi_{1}}$, $\mu_{2}=i\sqrt{1-\alpha}=\left\vert \mu
_{2}\right\vert e^{i\phi_{2}}$, and where the unitary matrix $R\left(
\phi\right)  $ is given by%
\begin{equation}
R\left(  \phi\right)  =\left(
\begin{array}
[c]{cc}%
\frac{e^{i\phi}}{\sqrt{2}} & \frac{e^{i\phi}}{\sqrt{2}}\\
\frac{e^{-i\phi}}{\sqrt{2}} & -\frac{e^{-i\phi}}{\sqrt{2}}%
\end{array}
\right)  \;,
\end{equation}
and thus in steady state [see Eq. (\ref{v P pm DLS}), recall that
$W_{\mathrm{T}1}=\left\vert W_{\mathrm{T}1}\right\vert e^{i\phi_{\mathrm{T}}}%
$, and note that $\left\vert \mu_{1}\right\vert ^{2}-\left\vert \mu
_{2}\right\vert ^{2}=\alpha+\left(  \delta-4\sqrt{D}z\right)  ^{2}$]%
\begin{equation}
\left(
\begin{array}
[c]{c}%
P_{+}\\
P_{-}%
\end{array}
\right)  =\frac{izP_{z0}T_{2}\left\vert W_{\mathrm{T}1}\right\vert
}{\left\vert \mu_{1}\right\vert ^{2}-\left\vert \mu_{2}\right\vert ^{2}%
}\left(
\begin{array}
[c]{c}%
\left\vert \mu_{1}\right\vert e^{-i\left(  \phi_{\mathrm{T}}+\phi_{1}\right)
}+\left\vert \mu_{2}\right\vert e^{i\left(  \phi_{\mathrm{T}}+\phi_{2}\right)
}\\
-\left\vert \mu_{1}\right\vert e^{i\left(  \phi_{\mathrm{T}}+\phi_{1}\right)
}-\left\vert \mu_{2}\right\vert e^{-i\left(  \phi_{\mathrm{T}}+\phi
_{2}\right)  }%
\end{array}
\right)  \;.
\end{equation}
The following holds%
\begin{equation}
P_{+}P_{-}=\left(  \frac{zP_{z0}T_{2}\left\vert W_{\mathrm{T}1}\right\vert
}{\left\vert \mu_{1}\right\vert \left(  1-\left\vert \frac{\mu_{2}}{\mu_{1}%
}\right\vert ^{2}\right)  }\right)  ^{2}g_{\mathrm{P}}\left(  \phi
_{\mathrm{T}}\right)  \;,
\end{equation}
where the phase--dependent gain $g_{\mathrm{P}}\left(  \phi_{\mathrm{T}%
}\right)  \ $is given by%
\begin{align}
g_{\mathrm{P}}  &  =\left\vert 1+\frac{\left\vert \mu_{2}\right\vert
e^{2i\left(  \phi_{\mathrm{T}}+\frac{\phi_{1}+\phi_{2}}{2}\right)  }%
}{\left\vert \mu_{1}\right\vert }\right\vert ^{2}\nonumber\\
&  =\left(  1+\frac{\left\vert \mu_{2}\right\vert }{\left\vert \mu
_{1}\right\vert }\right)  ^{2}\cos^{2}\left(  \phi_{\mathrm{T}}+\frac{\phi
_{1}+\phi_{2}}{2}\right)  +\left(  1-\frac{\left\vert \mu_{2}\right\vert
}{\left\vert \mu_{1}\right\vert }\right)  ^{2}\sin^{2}\left(  \phi
_{\mathrm{T}}+\frac{\phi_{1}+\phi_{2}}{2}\right) \nonumber\\
&  =1+\frac{2\left\vert \mu_{2}\right\vert }{\left\vert \mu_{1}\right\vert
}\cos\left(  2\phi_{\mathrm{T}}+\phi_{1}+\phi_{2}\right)  +\frac{\left\vert
\mu_{2}\right\vert ^{2}}{\left\vert \mu_{1}\right\vert ^{2}}\;.\nonumber\\
&  \label{f_P DLS}%
\end{align}
Note that as a function of $\phi_{\mathrm{T}}$ the gain $g_{\mathrm{P}}\left(
\phi_{\mathrm{T}}\right)  $ oscillates between a minimum value given by
$\left(  1-\left\vert \mu_{2}\right\vert /\left\vert \mu_{1}\right\vert
\right)  ^{2}$ and a maximum value given by $\left(  1+\left\vert \mu
_{2}\right\vert /\left\vert \mu_{1}\right\vert \right)  ^{2}$.

\section{Bosonization}

\label{SM_MM}

The method of Bosonization is widely employed to study spin
systems. As is discussed below, by implementing this method, the equations of
motion become nonlinear, and consequently, multistabilities, which are
otherwise excluded, become allowed. In the Holstein--Primakoff
transformation \cite{SM_Holstein_1098}, the operators $S_{\pm}$ and $S_{z}$
are expressed as $S_{+}=2B^{\dag}\left(  L-B^{\dag}B^{{}}\right)  ^{1/2}$,
$S_{-}=2\left(  L-B^{\dag}B^{{}}\right)  ^{1/2}B^{{}}$ and $S_{z}=-L+2B^{\dag
}B^{{}}$, where $L$ is the total number of spins, and where $B^{\dag}B$ is a
number operator. Self--consistency with the originally--assumed commutation
relations $\left[  S_{+},S_{-}\right]  =4S_{z}$ and $\left[  S_{z},S_{\pm
}\right]  =\pm2S_{\pm}$ is obtained by requiring that the operators $B^{{}}$
and $B^{\dag}$\ satisfy the Bosonic commutation relation $\left[  B,B^{\dag
}\right]  =1$.

The Hamiltonian $\mathcal{H}$ [see Eq. (\ref{H DLS}) in the main text] can be
expressed in terms of the operators $B^{{}}$ and $B^{\dag}$
\cite{SM_Hill_S227,SM_Wang_224410,SM_Zhang_987511,SM_Mathai_67001}. The
following holds $\left(  S_{+}S_{-}+S_{-}S_{+}\right)  /8=\left(  L-1\right)
B^{\dag}B^{{}}-B^{\dag}B^{\dag}B^{{}}B^{{}}+L/2$ [see the term proportional to
$\omega_{\mathrm{K}}$ in the Hamiltonian $\mathcal{H}$ given by Eq.
(\ref{H DLS}) in the main text]. The approximations $S_{+}\simeq
2L^{1/2}B^{\dag}$ and $S_{-}\simeq2L^{1/2}B^{{}}$, which are based on the
assumption that $\left\langle B^{\dag}B^{{}}\right\rangle /L\ll1$, yield
$S_{+}^{2}+S_{-}^{2}=4L\left(  B^{\dag}B^{\dag}+B^{{}}B^{{}}\right)  $ [see
the term proportional to $\omega_{\mathrm{A}}$ in the Hamiltonian
$\mathcal{H}$ given by Eq. (\ref{H DLS}) in the main text].

The terms proportional to $\Omega_{\mathrm{T}1}^{{}}$ and $\Omega
_{\mathrm{T}1}^{\ast}$ in the Hamiltonian $\mathcal{H}$ given by Eq.
(\ref{H DLS}) in the main text represent transverse driving. In the
Bosonization method, these terms are excluded from the closed--system
Hamiltonian $\mathcal{H}$, and transverse driving is instead accounted for by
introducing an external feedline \cite{SM_Gardiner_3761}. The coupling between
the feedline and the spins, which is assumed to be linear, is characterized by
a rate denoted by $\gamma_{1}$. Closed--system Heisenberg equations of motion
for the operators $B^{{}}$ and $B^{\dag}$ are thus derived by ignoring the
transverse driving terms in $\mathcal{H}$ (the approximation $L-1\simeq L$ is
implemented)%
\begin{align}
\frac{\mathrm{d}}{\mathrm{d}t}\left(
\begin{array}
[c]{c}%
B^{\dag}\\
B^{{}}%
\end{array}
\right)   &  =\left(
\begin{array}
[c]{cc}%
-i\omega_{z}+iL\omega_{\mathrm{K}} & iL\omega_{\mathrm{A}}\\
-iL\omega_{\mathrm{A}} & i\omega_{z}-iL\omega_{\mathrm{K}}%
\end{array}
\right)  \left(
\begin{array}
[c]{c}%
B^{\dag}\\
B^{{}}%
\end{array}
\right) \nonumber\\
&  +2i\omega_{\mathrm{K}}\left(
\begin{array}
[c]{c}%
-B^{\dag}B^{\dag}B^{{}}\\
B^{\dag}B^{{}}B^{{}}%
\end{array}
\right)  \;.\nonumber\\
&
\end{align}

The operator transformation [compare with Eq. (\ref{S pm T DLS}), and recall
that $X=\left(  1/2\right)  \left(  \sqrt{1+\omega_{\mathrm{A}}/\omega
_{\mathrm{K}}}+\sqrt{1-\omega_{\mathrm{A}}/\omega_{\mathrm{K}}}\right)
=1+O\left(  \omega_{\mathrm{A}}^{2}\right)  $ and $Y=\left(  1/2\right)
\left(  \sqrt{1+\omega_{\mathrm{A}}/\omega_{\mathrm{K}}}-\sqrt{1-\omega
_{\mathrm{A}}/\omega_{\mathrm{K}}}\right)  =\left(  1/2\right)  \omega
_{\mathrm{A}}/\omega_{\mathrm{K}}+O\left(  \omega_{\mathrm{A}}^{2}\right)  $]%
\begin{equation}
\left(
\begin{array}
[c]{c}%
\mathcal{B}_{+}\\
\mathcal{B}_{-}%
\end{array}
\right)  =\left(
\begin{array}
[c]{cc}%
e^{i\omega_{\mathrm{T}}t} & 0\\
0 & e^{-i\omega_{\mathrm{T}}t}%
\end{array}
\right)  \left(
\begin{array}
[c]{cc}%
X & Y\\
Y & X
\end{array}
\right)  \left(
\begin{array}
[c]{c}%
B^{\dag}\\
B^{{}}%
\end{array}
\right)  \;, \label{B pm T}%
\end{equation}
yields [recall that $X^{2}+Y^{2}=1$, $2XY=\omega_{\mathrm{A}}/\omega
_{\mathrm{K}}$, and $X^{2}-Y^{2}=\sqrt{1-\left(  \omega_{\mathrm{A}}%
/\omega_{\mathrm{K}}\right)  ^{2}}=1/\varrho$, where $\varrho=1/\sqrt
{1-\left(  \omega_{\mathrm{A}}/\omega_{\mathrm{K}}\right)  ^{2}}$]%
\begin{align}
&  \frac{\mathrm{d}}{\mathrm{d}t}\left(
\begin{array}
[c]{c}%
\mathcal{B}_{+}\\
\mathcal{B}_{-}%
\end{array}
\right) \nonumber\\
&  =i\left(
\begin{array}
[c]{cc}%
\omega_{\mathrm{T}}-\varrho\omega_{z}+\frac{L\omega_{\mathrm{K}}}{\varrho} &
\frac{\varrho\omega_{\mathrm{A}}\omega_{z}e^{2i\omega_{\mathrm{T}}t}}%
{\omega_{\mathrm{K}}}\\
-\frac{\varrho\omega_{\mathrm{A}}\omega_{z}e^{-2i\omega_{\mathrm{T}}t}}%
{\omega_{\mathrm{K}}} & -\omega_{\mathrm{T}}+\varrho\omega_{z}-\frac
{L\omega_{\mathrm{K}}}{\varrho}%
\end{array}
\right)  \left(
\begin{array}
[c]{c}%
\mathcal{B}_{+}\\
\mathcal{B}_{-}%
\end{array}
\right) \\
&  +2i\omega_{\mathrm{K}}\left(
\begin{array}
[c]{cc}%
e^{i\omega_{\mathrm{T}}t} & 0\\
0 & e^{-i\omega_{\mathrm{T}}t}%
\end{array}
\right)  \left(
\begin{array}
[c]{cc}%
X & Y\\
Y & X
\end{array}
\right)  \left(
\begin{array}
[c]{c}%
-B^{\dag}B^{\dag}B^{{}}\\
B^{\dag}B^{{}}B^{{}}%
\end{array}
\right)  \;.\nonumber\\
&
\end{align}
The transformation inverse to (\ref{B pm T}) yields in the RWA (recall that
$X^{2}+Y^{2}=1$)%
\begin{equation}
\left(
\begin{array}
[c]{cc}%
e^{i\omega_{\mathrm{T}}t} & 0\\
0 & e^{-i\omega_{\mathrm{T}}t}%
\end{array}
\right)  \left(
\begin{array}
[c]{cc}%
X & Y\\
Y & X
\end{array}
\right)  \left(
\begin{array}
[c]{c}%
-B^{\dag}B^{\dag}B^{{}}\\
B^{\dag}B^{{}}B^{{}}%
\end{array}
\right)  =\left(
\begin{array}
[c]{c}%
-\frac{X^{2}\mathcal{B}_{+}\mathcal{B}_{+}\mathcal{B}_{-}+2X^{2}%
Y^{2}\mathcal{B}_{+}\mathcal{B}_{-}\mathcal{B}_{+}+Y^{2}\mathcal{B}%
_{-}\mathcal{B}_{+}\mathcal{B}_{+}}{\left(  X^{2}-Y^{2}\right)  ^{3}}\\
\frac{X^{2}\mathcal{B}_{+}\mathcal{B}_{-}\mathcal{B}_{-}+2X^{2}Y^{2}%
\mathcal{B}_{-}\mathcal{B}_{+}\mathcal{B}_{-}+Y^{2}\mathcal{B}_{-}%
\mathcal{B}_{-}\mathcal{B}_{+}}{\left(  X^{2}-Y^{2}\right)  ^{3}}%
\end{array}
\right)  \;,
\end{equation}
and thus in this approximation [recall that $\omega_{z}=\omega_{0}%
+\Omega_{\mathrm{L}1}^{{}}\cos\left(  2\left(  \omega_{\mathrm{T}}%
+\omega_{\mathrm{f}}\right)  t\right)  $, and that $W_{\mathrm{A}}=\left(
\varrho/2\right)  \left(  \omega_{\mathrm{A}}/\omega_{\mathrm{K}}\right)
\Omega_{\mathrm{L}1}^{{}}$]%
\begin{align}
&  \frac{\mathrm{d}}{\mathrm{d}t}\left(
\begin{array}
[c]{c}%
\mathcal{B}_{+}\\
\mathcal{B}_{-}%
\end{array}
\right) \nonumber\\
&  =i\left(
\begin{array}
[c]{cc}%
\omega_{\mathrm{T}}-\varrho\omega_{z}+\frac{L\omega_{\mathrm{K}}}{\varrho} &
W_{\mathrm{A}}e^{-2i\omega_{\mathrm{f}}t}\\
-W_{\mathrm{A}}e^{2i\omega_{\mathrm{f}}t} & -\omega_{\mathrm{T}}+\varrho
\omega_{z}-\frac{L\omega_{\mathrm{K}}}{\varrho}%
\end{array}
\right)  \left(
\begin{array}
[c]{c}%
\mathcal{B}_{+}\\
\mathcal{B}_{-}%
\end{array}
\right) \nonumber\\
&  +2i\omega_{\mathrm{K}}\left(
\begin{array}
[c]{c}%
-\frac{X^{2}\mathcal{B}_{+}\mathcal{B}_{+}\mathcal{B}_{-}+2X^{2}%
Y^{2}\mathcal{B}_{+}\mathcal{B}_{-}\mathcal{B}_{+}+Y^{2}\mathcal{B}%
_{-}\mathcal{B}_{+}\mathcal{B}_{+}}{\left(  X^{2}-Y^{2}\right)  ^{3}}\\
\frac{X^{2}\mathcal{B}_{+}\mathcal{B}_{-}\mathcal{B}_{-}+2X^{2}Y^{2}%
\mathcal{B}_{-}\mathcal{B}_{+}\mathcal{B}_{-}+Y^{2}\mathcal{B}_{-}%
\mathcal{B}_{-}\mathcal{B}_{+}}{\left(  X^{2}-Y^{2}\right)  ^{3}}%
\end{array}
\right)  \;.\nonumber\\
&  \label{B pm T BBB}%
\end{align}

Damping is characterized by a linear rate given by $\gamma=\gamma_{1}%
+\gamma_{2}$, and a cubic nonlinear rate denoted by $\gamma_{3}$. The rate
$\gamma_{2}$ represents intrinsic FMSR damping (recall that $\gamma_{1}$
characterizes coupling to the feedline). Transverse driving, which is
delivered via the external feedline \cite{SM_Yurke_5054}, is characterized by
amplitude $\omega_{\mathrm{T}1}$ (in units of rate) and relative phase
$\phi_{\mathrm{T}}$.

Equations of motion for the expectation values $\beta_{\pm
}=\left\langle \mathcal{B}_{\pm}\right\rangle $ are obtained by applying
averaging to the operator equations of motion (\ref{B pm T BBB}), and by
adding terms representing damping and transverse driving. In the next step,
the mean field approximation is implemented by replacing terms having the form
$\left\langle \mathcal{B}_{\sigma_{1}}\mathcal{B}_{\sigma_{2}}\mathcal{B}%
_{\sigma_{3}}\right\rangle $, by $\left\langle \mathcal{B}_{\sigma_{1}%
}\right\rangle \left\langle \mathcal{B}_{\sigma_{2}}\right\rangle \left\langle
\mathcal{B}_{\sigma_{3}}\right\rangle $, where $\sigma_{n}\in\left\{
+,-\right\}  $ and $n\in\left\{  1,2,3\right\}  $. The validity of the mean
field approximation is arguably questionable, because, as can be seen from the
derivation below, terms having the form $\left\langle \mathcal{B}_{\sigma_{1}%
}\right\rangle \left\langle \mathcal{B}_{\sigma_{2}}\right\rangle \left\langle
\mathcal{B}_{\sigma_{3}}\right\rangle $ give rise to nonlinear dynamics.
Consequently, multistabilities, which are originally excluded, become allowed
when the mean field approximation is being implemented. The current manuscript
is mainly motivated by the difficulty to justify the mean field approximation.

The equations of motion for the expectation values $\beta_{\pm}=\left\langle
\mathcal{B}_{\pm}\right\rangle $ are (recall that $X^{2}+Y^{2}=1$)%
\begin{equation}
\frac{\mathrm{d}}{\mathrm{d}t}\left(
\begin{array}
[c]{c}%
\beta_{+}\\
\beta_{-}%
\end{array}
\right)  =M_{\beta}\left(
\begin{array}
[c]{c}%
\beta_{+}\\
\beta_{-}%
\end{array}
\right)  +\sqrt{2\gamma_{1}\omega_{\mathrm{T}1}}\left(
\begin{array}
[c]{c}%
ie^{-i\phi_{\mathrm{T}}}\\
-ie^{i\phi_{\mathrm{T}}}%
\end{array}
\right)  \;, \label{d/dt beta_pm}%
\end{equation}
where the $2\times2$ matrix $M_{\beta}$ is given by%
\begin{equation}
M_{\beta}=\left(
\begin{array}
[c]{cc}%
W_{\mathrm{b}}^{{}} & iW_{\mathrm{A}}e^{-2i\omega_{\mathrm{f}}t}\\
-iW_{\mathrm{A}}e^{2i\omega_{\mathrm{f}}t} & W_{\mathrm{b}}^{\ast}%
\end{array}
\right)  \;, \label{M_beta}%
\end{equation}
and $W_{\mathrm{b}}=i\left(  \omega_{\mathrm{T}}-\varrho\omega_{0}%
+L\omega_{\mathrm{K}}/\varrho-2\varrho^{3}\omega_{\mathrm{K}}\left(  1+\left(
\omega_{\mathrm{A}}/\omega_{\mathrm{K}}\right)  ^{2}/2\right)  \beta_{+}%
\beta_{-}\right)  -\gamma-\gamma_{3}\beta_{+}\beta_{-}$[compare with Eq.
(\ref{M_T mu12})]. Note that $W_{\mathrm{b}}=i\left(  \Omega_{\mathrm{d}%
}-2\omega_{\mathrm{K}}\beta_{+}\beta_{-}\right)  -\gamma-\gamma_{3}\beta
_{+}\beta_{-}+O\left(  \omega_{\mathrm{A}}^{2}\right)  $, where $\Omega
_{\mathrm{d}}=\omega_{\mathrm{T}}-\omega_{0}+L\omega_{\mathrm{K}}$.

For the case $\omega_{\mathrm{f}}=0$, steady state solution of Eq.
(\ref{d/dt beta_pm}) is given by%
\begin{align}
\left(
\begin{array}
[c]{c}%
\beta_{+}\\
\beta_{-}%
\end{array}
\right)   &  =-\sqrt{2\gamma_{1}\omega_{\mathrm{T}1}}\left(
\begin{array}
[c]{cc}%
\frac{W_{\mathrm{b}}^{\ast}}{\left\vert W_{\mathrm{b}}\right\vert
^{2}-W_{\mathrm{A}}^{2}} & -\frac{iW_{\mathrm{A}}}{\left\vert W_{\mathrm{b}%
}\right\vert ^{2}-W_{\mathrm{A}}^{2}}\\
\frac{iW_{\mathrm{A}}}{\left\vert W_{\mathrm{b}}\right\vert ^{2}%
-W_{\mathrm{A}}^{2}} & \frac{W_{\mathrm{b}}^{{}}}{\left\vert W_{\mathrm{b}%
}\right\vert ^{2}-W_{\mathrm{A}}^{2}}%
\end{array}
\right)  \left(
\begin{array}
[c]{c}%
ie^{-i\phi_{\mathrm{T}}}\\
-ie^{i\phi_{\mathrm{T}}}%
\end{array}
\right) \nonumber\\
&  =-\frac{\sqrt{2\gamma_{1}\omega_{\mathrm{T}1}}}{\left\vert W_{\mathrm{b}%
}\right\vert ^{2}-W_{\mathrm{A}}^{2}}\left(
\begin{array}
[c]{c}%
iW_{\mathrm{b}}^{\ast}e^{-i\phi_{\mathrm{T}}}-W_{\mathrm{A}}e^{i\phi
_{\mathrm{T}}}\\
-W_{\mathrm{A}}e^{-i\phi_{\mathrm{T}}}-iW_{\mathrm{b}}^{{}}e^{i\phi
_{\mathrm{T}}}%
\end{array}
\right)  \;,\nonumber\\
&  \label{beta pm SS MM}%
\end{align}
and thus the steady state value of $E\equiv\left\vert \beta_{\pm}\right\vert
^{2}=\beta_{+}\beta_{-}$ can be calculated by solving the equation (note that
$W_{\mathrm{b}}$\ depends on $E$)%
\begin{equation}
E=\frac{2\gamma_{1}\omega_{\mathrm{T}1}g_{\mathrm{M}}}{\left\vert
W_{\mathrm{b}}\right\vert ^{2}\left(  1-\left\vert \frac{W_{\mathrm{A}}%
}{W_{\mathrm{b}}}\right\vert ^{2}\right)  ^{2}}\;, \label{E_beta}%
\end{equation}
where the phase--dependent gain $g_{\mathrm{M}}\left(  \phi_{\mathrm{T}%
}\right)  $\ is given by [compare to Eq. (\ref{f_P DLS})]%
\begin{equation}
g_{\mathrm{M}}=\left\vert 1+\frac{iW_{\mathrm{A}}e^{2i\phi_{\mathrm{T}}}%
}{W_{\mathrm{b}}^{\ast}}\right\vert ^{2}\;.
\end{equation}
The notation $iW_{\mathrm{A}}/W_{\mathrm{b}}^{\ast}=\eta_{\mathrm{A}}%
e^{i\phi_{\mathrm{A}}}$, where both $\eta_{\mathrm{A}}$ and $\phi_{\mathrm{A}%
}$ are real, allows expressing the gain $g_{\mathrm{M}}$ as%
\begin{align}
g_{\mathrm{M}}  &  =\left(  1+\eta_{\mathrm{A}}\right)  ^{2}\cos^{2}\left(
\phi_{\mathrm{T}}+\frac{\phi_{\mathrm{A}}}{2}\right)  +\left(  1-\eta
_{\mathrm{A}}\right)  ^{2}\sin^{2}\left(  \phi_{\mathrm{T}}+\frac
{\phi_{\mathrm{A}}}{2}\right) \nonumber\\
&  =1+2\eta_{\mathrm{A}}\cos\left(  2\phi_{\mathrm{T}}+\phi_{\mathrm{A}%
}\right)  +\eta_{\mathrm{A}}^{2}\;.\nonumber\\
&  \label{g_M}%
\end{align}
Similarly to the rapid disentanglement model, the steady state response is
periodic in the phase $\phi_{\mathrm{T}}$ with period $\pi$ [compare to Eq.
(\ref{f_P DLS})].

Consider the case where both $\omega_{\mathrm{A}}$ and $\beta_{+}\beta_{-}$
are assumed to be sufficiently small to validate an approximation, in which
terms of order $O\left(  \omega_{\mathrm{A}}^{2}\right)  $ are disregarded,
and terms of first order in $\omega_{\mathrm{A}}$ are evaluated in the limit
$\beta_{+}\beta_{-}\rightarrow0$. For this case%
\begin{equation}
\eta_{\mathrm{A}}=\left\vert \frac{W_{\mathrm{A}}}{W_{\mathrm{b}}^{\ast}%
}\right\vert =\frac{\omega_{\mathrm{A}}\Omega_{\mathrm{L}1}^{{}}}%
{2\omega_{\mathrm{K}}\sqrt{\Omega_{\mathrm{d}}^{2}+\gamma^{2}}}\;,
\label{eta_A}%
\end{equation}
and Eq. (\ref{E_beta}) yields a cubic polynomial equation for $E$ given by Eq.
(\ref{E MM SS}) of the main text.

\begin{figure}[ptb]
\begin{center}
\includegraphics[width=3.0in,keepaspectratio]{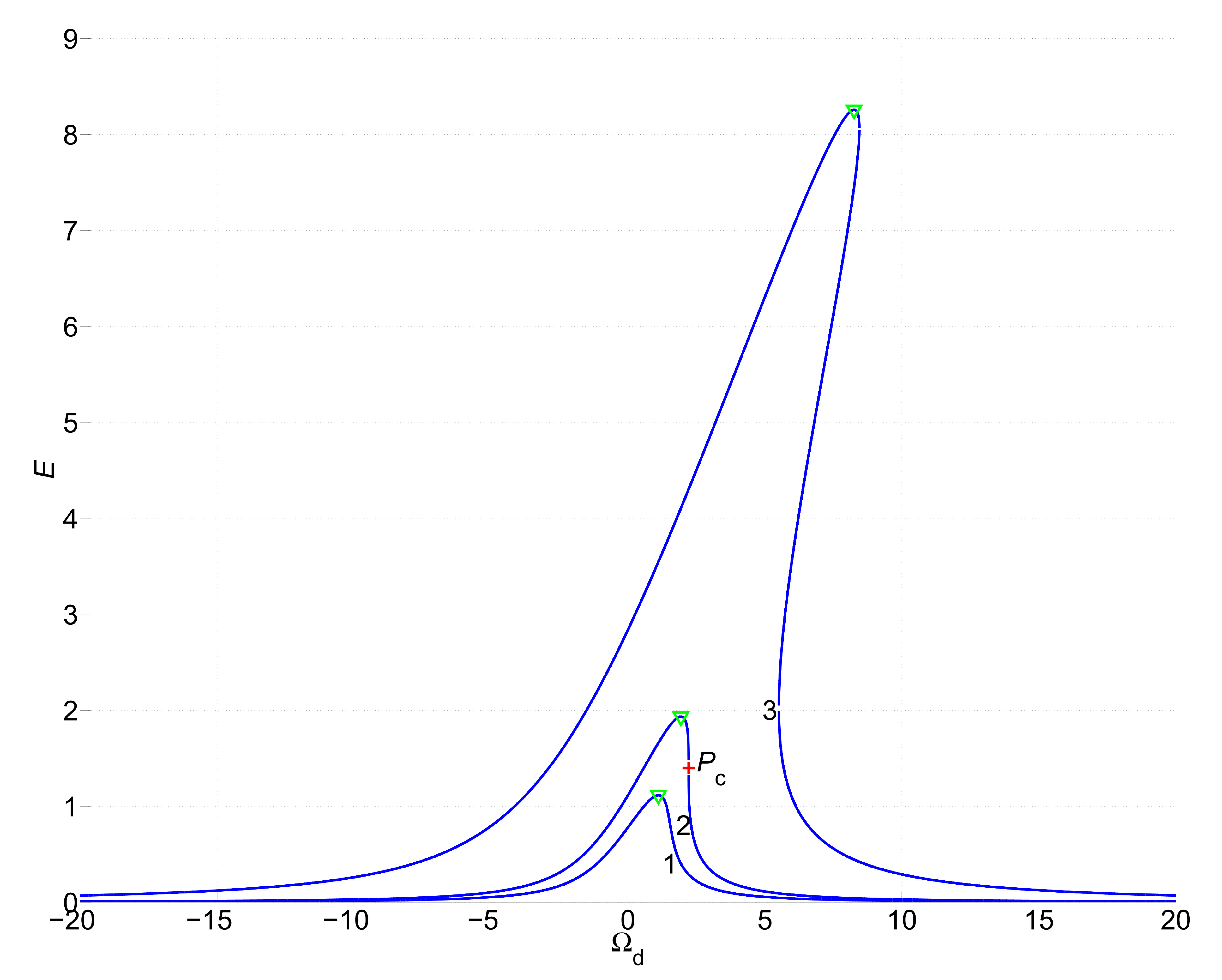}
\end{center}
\caption{{}Bosonization. The steady state value of magnon number expectation
value $E$ is calculated as a function of the detuning angular frequency
$\Omega_{\mathrm{d}}$ by solving the cubic polynomial equation given by Eq.
(\ref{E MM SS}) of the main text. Assumed parameters' values are $\gamma
_{1}/\gamma=0.5$, $\gamma_{3}/\gamma=0.1$ and $\omega_{\mathrm{K}}/\gamma=1$.
The driving amplitudes $\Omega_{1}$ for the curves labelled by the integers 1,
2 and 3 are $0.5\Omega_{\mathrm{1c}}$, $\Omega_{\mathrm{1c}}$ and
$10\Omega_{\mathrm{1c}}$, respectively, where $\Omega_{\mathrm{1c}}$ is the
value of $\Omega_{\mathrm{1}}$ at the bistability onset point, which is
labeled by a red cross symbol. Peak points are labeled by green triangles.}%
\label{FigME}%
\end{figure}

For any given transverse driving amplitude $\Omega_{1}\equiv\omega
_{\mathrm{T}1}g_{\mathrm{M}}$, the cubic polynomial equation allows
calculating the magnon number expectation value $E$ as a function of the
angular frequency detuning $\Omega_{\mathrm{d}}$. The plots shown in Fig.
\ref{FigME} demonstrate the transition from the mono--stability $\Omega
_{\mathrm{1}}<\Omega_{\mathrm{1c}}$ to the bistability $\Omega_{\mathrm{1}%
}>\Omega_{\mathrm{1c}}$ regions, where $\Omega_{\mathrm{1c}}$ is the value of
$\Omega_{\mathrm{1}}$ at the bistability onset point (see the red cross symbol
in Fig. \ref{FigME}) \cite{SM_Yurke_5054}.

Note that the analysis above is based on the assumption that magnetization is
uniform. This simplifying assumption has been employed before to model Kerr
nonlinearity in FMSR \cite{SM_Wang_224410}. For the more general case, the
response of the magnetic medium is decomposed into spin waves
\cite{SM_Suhl_209}. Note, however, that spin wave theory
\cite{SM_Stancil_Spin} is based on the same Bosonization method (i.e.
Holstein--Primakoff transformation \cite{SM_Holstein_1098}) that is employed
in this section. As was shown above in section \ref{SM_EOM}, the Heisenberg
equations of motion that are derived from the system's Hamiltonian are all
linear. Multistability, which is excluded by this linearity, becomes possible
when Bosonization is implemented, regardless of whether or not magnetization
is assumed to be uniform.

What is the validity range of the assumption that magnetization is uniform?
The system's response generally depends on the dispersion relation of spin
waves. The effect of exchange interaction on the dispersion relation is
characterized by the dimensionless parameter $\epsilon_{\mathrm{ex}}%
=\omega_{\mathrm{M}}\lambda_{\mathrm{ex}}k^{2}/\omega_{0}$, where
$\omega_{\mathrm{M}}=-\gamma_{\mathrm{e}}\mu_{0}M_{\mathrm{S}}$,
$\gamma_{\mathrm{e}}/2\pi=28\operatorname{GHz}\operatorname{T}^{-1}$ is the
gyromagnetic ratio, $\mu_{0}$ is the free space permeability, $M_{\mathrm{S}}$
is the saturated magnetization, $\lambda_{\mathrm{ex}}$ is the exchange
constant, $k=2\pi/\lambda_{\mathrm{m}}$, $\lambda_{\mathrm{m}}$\ is the spin
wavelength, and $\omega_{0}$ is the Larmor angular frequency [see Eq. (5.18)
of Ref. \cite{SM_Stancil_Spin}]. For sufficiently small values of the
transverse $\omega_{\mathrm{d}}$ and longitudinal $\omega_{\mathrm{f}}$
detuning angular frequencies, the effect of exchange--induced dispersion can
be disregarded \cite{Rezende_893}. From the condition $\epsilon_{\mathrm{ex}%
}\simeq1$ for the case $\lambda_{\mathrm{m}}=R_{\mathrm{s}}$ one finds that
exchange--induced dispersion can be approximately disregarded provided that
$\max\left(  \left\vert \omega_{\mathrm{d}}\right\vert ,\left\vert
\omega_{\mathrm{f}}\right\vert \right)  \lesssim\omega_{\mathrm{D}}$, where
$\omega_{\mathrm{D}}=\omega_{\mathrm{M}}\lambda_{\mathrm{ex}}\left(
2\pi/R_{\mathrm{s}}\right)  ^{2}$. For our experimental setup $\omega
_{\mathrm{D}}/\left(  2\pi\right)  \simeq2\operatorname{kHz}$
\cite{SM_Stancil_Spin}.

\section{Bistability}

\label{SM_BS}

\begin{figure}[ptb]
\begin{center}
\includegraphics[width=3.0in,keepaspectratio]{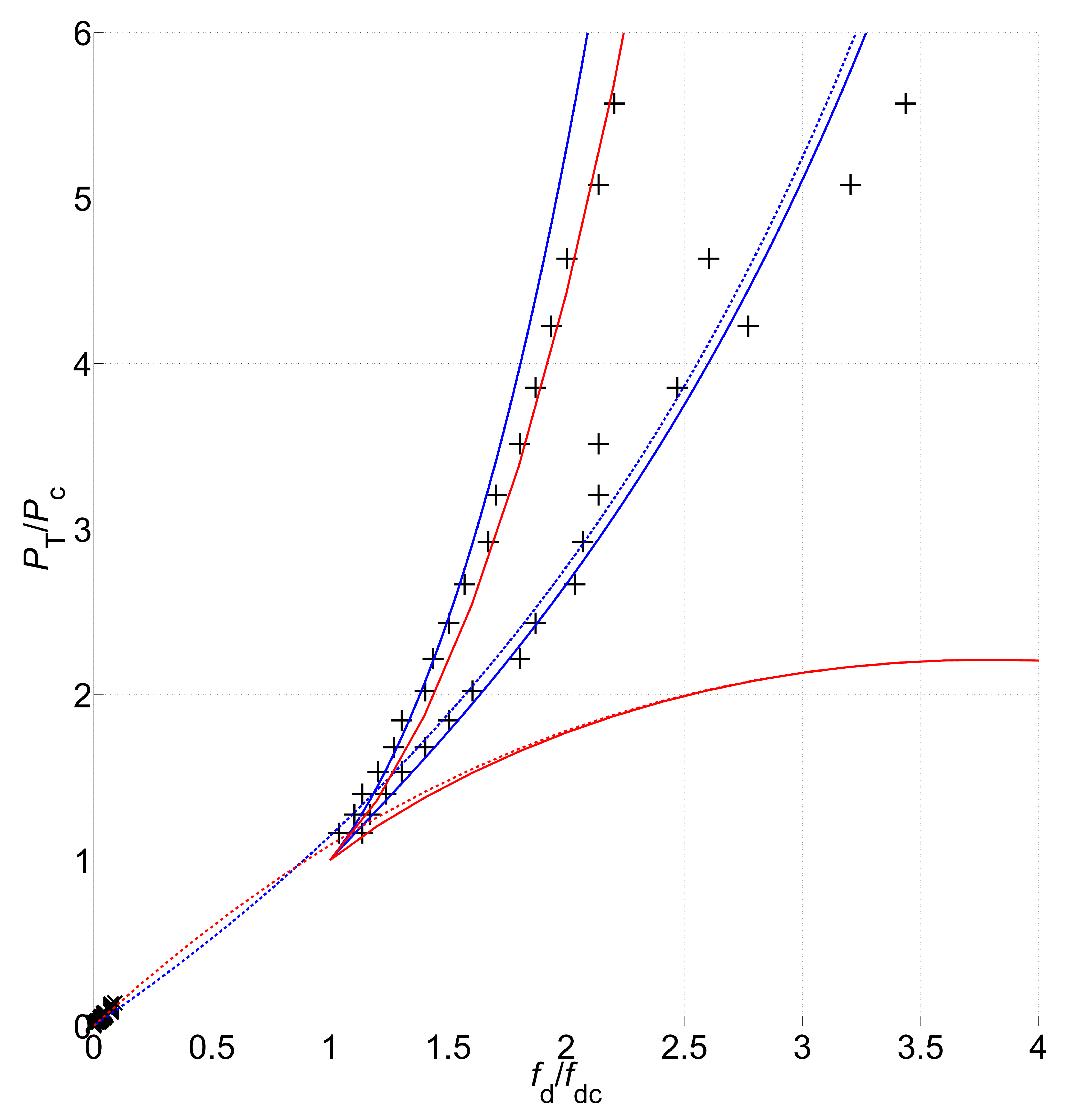}
\end{center}
\caption{{}Bistability. Onset of bistability occurs at driving frequency
detuning $f_{\mathrm{d}}$ denoted by $f_{\mathrm{dc}}$, and driving power
$P_{\mathrm{T}}$ denoted by $P_{\mathrm{c}}$. Measured jump and peak points
are labeled by the symbols $+$ and $\times$, respectively. Theoretical
predictions derived from the rapid disentanglement and Bosonization--based
models are colored by blue and red, respectively. The following measured FMSR
parameters are used for the calculations $\omega_{0}/\left(  2\pi\right)
=4.062101 \operatorname{GHz}$, $P_{\mathrm{c}}=13.5$ dBm, $\gamma_{1}%
/\gamma=0.4$ and $\gamma_{3}/\omega_{\mathrm{K}}=5.8\times10^{-2}$.}%
\label{FigSP}%
\end{figure}

Both the rapid disentanglement model (see section \ref{SM_RDM}) and the
Bosonization--based model (see section \ref{SM_MM}) predict bistability. The
plot in Fig. \ref{FigSP} presents a comparison between experimental mapping of
the region of bistability, which is measured with transverse driving only, and
theoretical predictions. The experimental mapping is performed by sweeping
both transverse driving detuning frequency $f_{\mathrm{d}}$ and transverse
driving power $P_{\mathrm{T}}$, and monitoring FMSR response. Above a critical
value of the driving power, which is denoted by $P_{\mathrm{c}}$, hysteresis
is observed in the dependency on $f_{\mathrm{d}}$. The driving detuning
frequency $f_{\mathrm{d}}$ at the bistability onset point (i.e. for
$P_{\mathrm{T}}=P_{\mathrm{c}}$) is denoted by $f_{\mathrm{dc}}$. For any
given value of transverse driving power $P_{\mathrm{T}}$ in the region of
bistability (i.e. for $P_{\mathrm{T}}>P_{\mathrm{c}}$), the lower (upper)
frequency bound of the region of bistability is determined by sweeping
$f_{\mathrm{d}}$ upwards (downwards), and identifying the driving detuning
frequency $f_{\mathrm{d}}$ at which the response exhibits a sharp jump. The
measured jump points are labelled in Fig. \ref{FigSP} using the symbol $+$.
The symbol $\times$ is used to label measured peak points.

Theoretical predictions derived from the rapid disentanglement model are blue
colored [see the cubic polynomial equation (\ref{z= DLS})], whereas the color
red is used for predictions derived from the Bosonization--based model [see
the cubic polynomial equation (\ref{E_beta})]. For both models, calculated
jump points are represented by solid lines, whereas dashed lines are used for
calculated peak points. The data--theory comparison presented in Fig.
\ref{FigSP} indicates that the rapid disentanglement model better aligns with
the experimental results.

\end{document}